\documentclass[12pt]{article}

\usepackage{scicite}

\usepackage{times}

\usepackage{graphicx} 
\usepackage{amsmath}
\usepackage{caption}
\usepackage{xurl}

\newenvironment{sciabstract}{%
\begin{quote} \bf}
{\end{quote}}

\title{Prognostic relevance of gene-expression signatures}

\author
{Dimitrij Tschodu,$^{1\ast}$ Jürgen Lippoldt,$^{1}$ Pablo Gottheil$^{1}$,\\ Anne-Sophie Wegscheider,$^{2}$ Axel Niendorf,$^{2}$ Josef A. Käs,$^{1}$\\
\normalsize{$^{1}$Peter Debye Institute, Leipzig University, Leipzig, Germany}\\
\normalsize{$^{2}$MVZ Prof. Dr. med. A. Niendorf Pathologie Hamburg-West GmbH,}\\ 
\normalsize{Institute for Histology, Cytology and Molecular Diagnostics, Hamburg, Germany}\\
\\
\normalsize{$^\ast$To whom correspondence should be addressed; E-mail:  dimitrij.tschodu@uni-leipzig.de.}\\
\\
\normalsize{\textbf{One sentence summary:}}\\
\normalsize{Prognostic gene-expression signatures are fundamentally limited in their prognostic power.}
}

\date{}

\begin{document}

\baselineskip24pt


\maketitle

\begin{sciabstract}
   		Cancer prognosis can be regarded as estimating the risk of future outcomes from multiple variables. In prognostic signatures, these variables represent expressions of genes that are summed up to calculate a risk score. However, it is a natural phenomenon in living systems that the whole is more than the sum of its parts.
   		We hypothesize that the prognostic power of signatures is fundamentally limited without incorporating emergent effects. Convergent evidence from a set of unprecedented size ($\approx$10,000 signatures) implicates a maximum prognostic power. We show that a signature can correctly discriminate patients’ prognoses in no more than 80\% of the time. Using a simple simulation, we show that more than 50\% of the potentially available information is still missing at this value. 
\end{sciabstract}

\subsection*{Introduction}

Clinicians assimilate a variety of data from the clinical history, pathological characteristics, and 
molecular biomarkers to guide decision-making. Especially biomarkers have increased 
clinicians' prognostic capabilities over the last decade \cite{Andre2011}.     
Such biomarkers represent measurable molecular characteristics that can be used to classify patients 
into subgroups: predictive biomarkers divide individuals into subgroups based on the likelihood
a patient responds to a drug, and prognostic biomarkers divide individuals into subgroups based on         
a patient's risk for a clinical endpoint such as death or metastasis \cite{nicolini2018prognostic}. 
Especially in breast cancer, biomarkers have exhibited great promise and provided some of the most
successful examples of translating knowledge from genomic studies to clinical applications \cite{nicolini2018prognostic}.   

A striking example of this translation are the so-called gene-expression signatures: 
biomarkers that are based solely on genes or combined groups of genes. 
Advancement in signature development was enhanced with the advent of high-throughput technologies \cite{Latha2020}
as well as with the simplicity of constructing a signature: most of the prognostic signatures have been constructed by 
simply summing up and normalizing weighted expression 
values of genes or groups of genes to produce a risk score, 
in a statistical process known as the Cox proportional hazards model \cite{cox2018analysis}. 
This simplicity has been recognized by the medical community \cite{Chibon2013a},
and several prognostic signatures have been clinically validated, resulting in the development of commercial 
tests such as EndoPredict\textregistered, MammaPrint\textcopyright, Oncotype DX\textregistered, or Prosigna\textregistered.  

Despite considerable interest and effort, a growing number of signatures proposed for clinical risk assessment has 
largely failed to surpass
the performance of standard clinical characteristics \cite{Poste2011, wegscheider2021altona}, such as the 
Nottingham Prognostic Index \cite{haybittle1982prognostic} used in breast cancer,
which combines the size and grade of the tumor as well as the number of nearby invaded lymph nodes.   
Current research has noticed slow progress in developing useful signatures relating it to inconsistent results \cite{Koscielny2010},
poor study design \cite{Michiels2005}, improper validation \cite{Simon2005}, and the finding that
even a random set of genes can be prognostic \cite{Venet2011}, as well as to the finding that the most 
prognostic signatures lack any biological meaning with respect to disease etiology \cite{Manjang2021}.

For the same reasons, however, one can assume that gene-expression signatures lack fundamental information
which cannot be captured by reducing the connections between the biology and systemic disease to a single set of genes.       
Such lack of fundamental information suggests to consider cancer as an emergent phenomenon which has irreducible properties 
that cannot be explained by a bottom up approach as the sum of its molecular changes.
In fact, several theoretical studies, simulations, and experimental observations point towards the dominant role of the 
(host) system over the cell in dictating the behavior of cells in cancer progression \cite{Barcellos-Hoff2008, Ducasse2015, Soto2005, Chang2013, Gatenby2002}.
Thus, we hypothesize that a lack of such fundamental information in gene-expression signatures leads to an inherent limit of their prognostic power.

Here, we perform a comprehensive analysis of prognostic signatures in breast cancer patients, determining their current limits and estimating
the potential maximum prognostic power. 
To quantify the prognostic power, we measure the concordance index which is a measure of how well patients 
with different prognoses can be discriminated.	
Three different approaches for selecting genes are considered to explore a wide range of different gene signatures:
random sampling, since it addresses the finding that even random signatures can be prognostic; 
collecting genes from signatures reported in the literature, because it addresses expert knowledge
and functionally related genes; and selecting genes
using the state-of-the-art machine learning methods, since ensemble algorithms such as random forests can 
detect nonlinear relations between genes (\textit{Materials and Methods}). These selections are performed on 8 established breast cancer 
data bases, which provide 
expression values quantified with different methods such as reverse transcription-PCR or DNA microarrays (\textit{Materials and Methods, Datasets}).
Based on these selections, we develop signatures with 9 different machine learning models. 

This work provides clinicians and researches – to the best of our knowledge – with the largest examination 
of the prognostic utility of gene-expression signatures ($\approx$10,000 signatures). Our results illustrate that it is unlikely to find a gene-expression signature
which would perfectly order patients' survival times in more than 80\% of the time; and 50\% of potential 
information is still hidden for such signature. 

\section*{Study design}

The concept of maximum prognostic power demands 
at least three assumptions to be satisfied. First, it is independent on the number of patients on which 
the prognosis is based, i.e. prognosis cannot be improved by simply collecting more data.  
Second, it is independent on the selection method of variables, i.e. of genes, 
on which the prognosis is based. Last, it is independent on the prognostic model, i.e. on the procedure or 
algorithm which implements the prediction of survival.

\begin{figure*}[tbhp]
	\centering
	\includegraphics[width=1.0\textwidth]{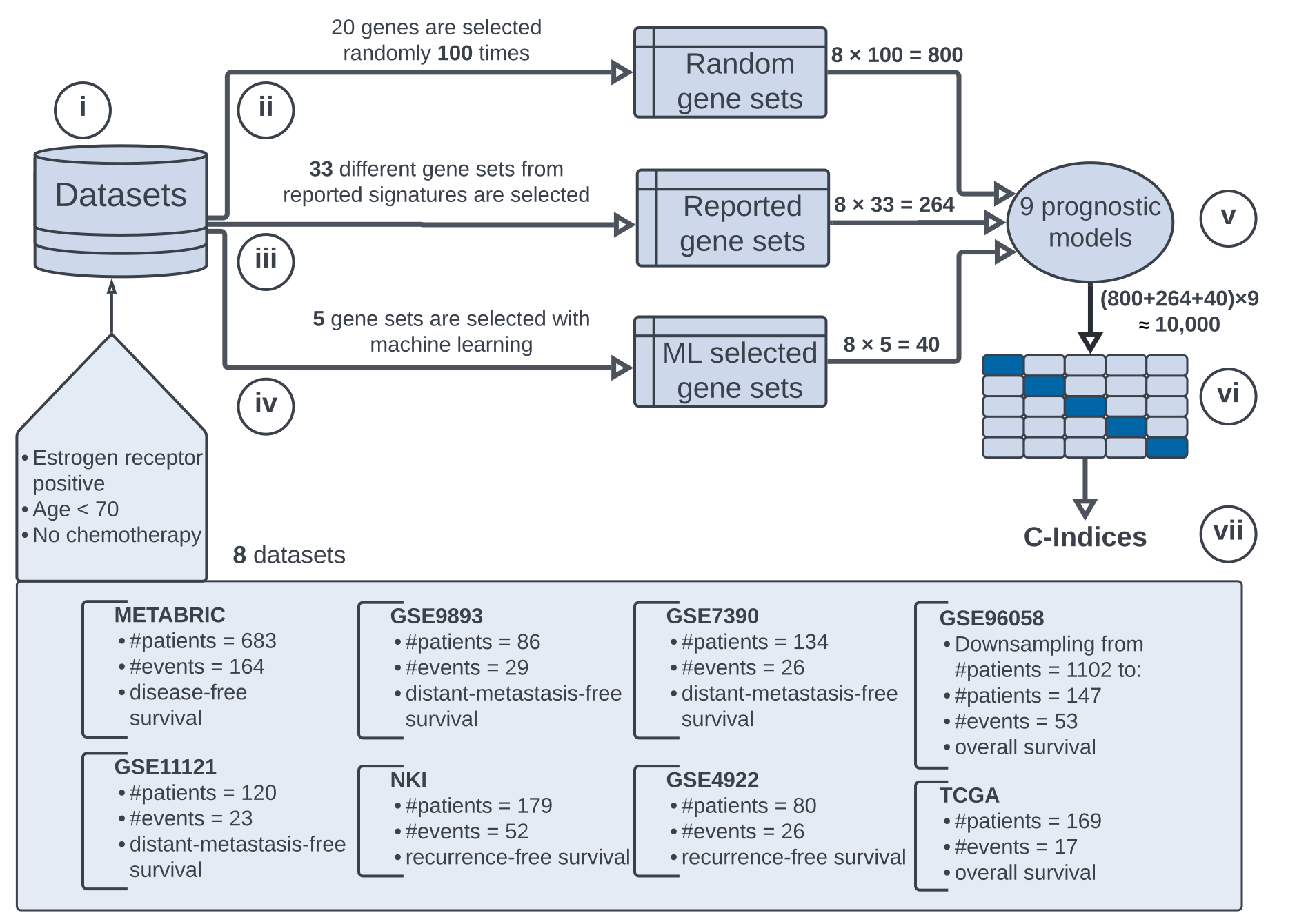}
	\caption{\textbf{Our 7-step analysis.} Steps are indicted in circles. i) We collect and filter 8 datasets with different number of patients containing expression data and information about survival (box at the bottom). Filtering is conducted by selecting estrogen-receptor positive patients under the age of 70 years who did not receive chemotherapy. We  use 3 approaches to select gene sets in each dataset: ii) 20 genes are sampled 100 times at random (Random gene sets), iii) 33 different
		gene sets are selected that were reported in the literature (Reported gene sets), iv) gene sets are selected with 5 machine learning methods (ML selected gene sets).
		Each gene set serves as input to a prognostic model. v) Overall 9 prognostic models are developed, resulting in $8\times(100+33+5)\times9 = 9936 \approx 10,000$ signatures. vi) Evaluation is performed using 5-fold cross-validation, whereby each dataset
		is randomly permuted and split 5 times, each time in 4 training and 1 test sets (the center right of the figure, dark blue indicates
		test set and light blue training set). vii) Prognostic power is measured by calculating the median C-index based on the 5 cross-validation test sets.}
	\label{fig:flow_diagram}
\end{figure*}

To provide an estimation of maximum prognostic power which attempts to justify these assumptions, we perform a 7-step analysis summarized
in Fig. \ref{fig:flow_diagram}:

\begin{itemize}
	\item[i] Datasets, i.e. expression values and patient's information about survival, are collected from 8 different sources.
	Stemming from different sources, these datasets vary in the number of patients and statistical properties such as primary end-point 
	(\textit{Materials and Methods}), which guarantees that dependency on the number of patients can be examined.
	
	\item[ii] We select 100 gene sets – a gene set is a list of genes that is used for prognosis – at random in each dataset,  
	resulting in $8\times100=800$ gene sets. We have determined the optimal number of randomly selected genes in a gene set 
	to be approximately 20 (\textit{Supplementary Random Signature Size}), which confirms the finding of Chou et. al. (Fig. 4 in \cite{chou2013gene}), who found that 20 genes is optimal by using neural networks.   
	The sampling of random signatures is based on recent studies that emphasize the role of random selections in prognosis \cite{Venet2011, Goh2018, Tschodu2022},
	whereby random signatures outperform published and known signatures.
	Goh et.al \cite{Goh2018} termed this phenomenon as random signature superiority and related it to the fact that
	random genes are inseparably correlated with proliferation genes, including genes involved in
	cell cycle, cell death, contact-based growth inhibition, and so forth.     
	
	\item[iii] 	Several gene-expression signatures have been reported in the literature and associated
	with clinical outcome, so that they are expected to provide higher prognostic power than random signatures.  
	We curate 33 gene sets from gene signatures reported in previous studies \cite{Huang2018} such as 
	OncotypeDx, EndoPredict, MammaPrint, or ProSigna, resulting in $8\times33 = 264 $ gene sets. These gene sets are  
	described in detail in \textit{SI Appendix, Supplementary Reported Gene Selections}.
	
	\item[iv] Gene sets are chosen with one standard selection method (UM, univariable model) that detects linear 
	relations between genes; and 4 machine learning methods that are based on so-called random forests which can 
	detect non-linearity between genes (\textit{Materials and Methods}), resulting in $8\times5 = 40 $ gene sets.
	
	\item[v] Prognostic models for each gene set are developed (\textit{Materials and Methods}).
	A prognostic model provides a statistical function that aims to predict the time from a fixed time point 
	to an event, such as the time from surgery to death, by modeling the relation between one or more 
	variables (genes) and a response (event). The inherent aspect of prognostic modeling is the presence 
	of censored data.   
	For example, a censoring occurs if a patient is lost to followup or the event does not occur within the study duration.
	We use 9 prognostic models that can handle censored data, resulting in overall $8\times(100+33+5)\times\times9 \approx 10,000 $ gene
	expression signatures. 
	The Cox proportional hazards model \cite{cox1972regression} is employed, since it is the most common method for analyzing censored data \cite{cox2018analysis}. 
	However, this model detects only linear effects between variables. In order to detect non-linear effects
	we use 8 machine learning models based on random forests and so-called gradient boosting machines \cite{freund1997decision, friedman2000additive} (\textit{Materials and Methods}).
	
	\item[vi] 
	Each prognostic model is evaluated using the 5-fold-cross-validation, since it can be applied to
	datasets of different sizes \cite{Picard1984}. 
	The $k$-fold-cross-validation is a resampling method
	that divides data into $k$ sets (folds) of approximately equal size \cite{hastie2009elements}. The model is trained on $(k-1)$ folds and the remaining fold is
	used as test set for computing the prognostic power. This procedure is repeated $k$ times while a different fold 
	is chosen each time as the test set. 
	
	\item[vii]
	The median concordance index, also called C-index \cite{Harrell1982}, is computed. 
	The C-index describes the ability of a prognostic model
	to separate patients with good and poor outcomes (\textit{Materials and Methods}). 
	C-index of 0.5 denotes a completely random prognosis 
	and a value of 1.0 implies that one can perfectly order 
	the predicted temporal survival probabilities of patients: a patient with a 
	higher survival time would get a higher probability than a patient with a shorter survival time. 
	A C-index = 0 describes the perfect anti-concordance, where the predicted survival probabilities 
	are inversely proportional to the survival times.  
	In his seminal work, Harrell \cite{Harrell1982} provides the interpretation of the C-index as the percentage of patients 
	that can be correctly ordered. For instance, a value of 0.7 indicates that one can correctly order patients' prognoses 
	70\% of all cases.
	
\end{itemize}

Steps ii-iv~guarantee that the dependency on the selection method can be examined; and step v~verifies that
the maximum prognostic power is not confounded by a prognostic model.

\section*{Random signature superiority in breast cancer}

We evaluated the prognostic power of gene-expression signatures generated at random, since
the so-called random signature superiority (RSS) is a known but still an underexplored area \cite{Goh2018}.
In order to select a gene set, 20 genes were sampled at random in each dataset.  
We selected 100 gene sets, developed a 
prognostic model, and measured the corresponding C-index for each gene set. Fig. \ref{fig:ridges_random}
shows the resulting C-indices.

Here, each data point represents the median C-index computed by the 5-fold cross-validation and based on a single random gene set.
Each row corresponds to a machine learning prognostic model used to compute the risk score.
There are 9 prognostic models of 100 random gene sets in each of 8 datasets, 
resulting in overall $9\times100\times8 = 7200$ data points.
On the left and on the right sides of each ridge plot, the median of the sample medians (MOM) and the maximum 
C-index (MAX) are shown, respectively. The median of the sample medians denotes the median value of medians
per prognostic model and can be interpreted as the center of the distribution.    
Additionally, Fig. \ref{fig:ridges_random} shows the density plots, which approximate
the distributions of the C-indices. 

A critical question is whether random signatures are suited to test a potential maximum prognostic power,
i.e. whether RSS applies to these data.  
The signature size is one of the major factors influencing
RSS \cite{Goh2018}. Thus, to investigate how frequently RSS
occurs, we calculated the number of random signatures performing above the C-index of 
the reported 26-gene signature (which has roughly the same size as random signatures, see \textit{Supplementary Reported Gene Selections}) for each prognostic model and averaged this value over all datasets.
We found (\textit{Supplementary Random Signature Superiority}) that more than 60\% of random signatures outperform the aforementioned reported signature in 4 of 8 datasets, exactly 49\% in one dataset, and less than 22\% in the remaining 3 datasets. Averaging across datasets, 
44\% of random signatures outperform the aforementioned reported signature.
These results demonstrate that RSS is strongly present in the context of breast cancer gene expression. Consequently,
they can be used to test if gene-expression signatures exhibit a maximum prognostic power.      

Next, we examined whether the prognostic power can be increased by collecting a larger number of patients. 
As shown in Fig. \ref{fig:ridges_random} the center of the distribution and the variability of C-indices differ 
across the datasets. 
For this, we investigated whether the MOM and the median absolute 
deviation (MAD) correlates with the number of patients 
as well as with the event rate in a dataset. The event rate is the ratio of the number of events to the number of patients
and represent a clinically relevant quantity, since prognostic quantities can vary by event rate \cite{Cook2018}.       
These dependencies are plotted in \textit{Supplementary Dataset Dependency} for each prognostic model along with the  
correlation coefficients and their p-values. As can be inspected there, the MOM and MAD seem to be uncorrelated 
with both the number of patients and with the event rate.   	
Thus, our data demonstrate that the overall prognostic power cannot be increased by collecting a larger number of patients.  

Fig. \ref{fig:ridges_random} shows also that the best performing prognostic model is different in each dataset.
Consequently, the values of maximum prognostic power are essentially unaffected by the choice of a prognostic model.                 

The points above demonstrate that random signatures can be used to compute the maximum prognostic power. 
Although the overall highest C-index is 0.84 (GSE11121, Rank-RF), Fig. \ref{fig:ridges_random} 
illustrates that the maximum prognostic power seems to be around the C-index of 0.8 for all models in all datasets.

\section*{Maximum prognostic power of current signatures}

More than 30 different gene-expression signatures have been reported so far \cite{Huang2018}. 
These signatures are expected to outperform random signatures, since they have been associated 
with clinical outcomes in the original studies.     

We adopted the approach described in \cite{Huang2018}, whose authors searched PubMed for
breast cancer signatures or classifiers and collected the gene lists from the original publications.
The majority of the corresponding gene sets (28 gene sets) has been used for prognosis, the rest
5 gene sets have been utilized for prediction, i.e. to predict response to a drug. 
We used these gene lists and the procedure described in \textit{Supplementary Reported Gene Selections}
to select the corresponding gene sets in each dataset. 

As can be seen in Fig. \ref{fig:ridges_signatures} , the C-indices are  
higher compared with the C-indices of random signatures. To quantify these differences,  
we show the distributions  
in form of the violin plots – i.e. box plots showing probability distributions –  for each model and each dataset in \textit{Supplementary Comparison Random And Reported Signatures}, 
and compare the distribution by using the Wilcoxon rank sum test, since the data are not normally distributed.
As can be seen, reported signatures tend indeed to have higher C-indices than random signatures, although the level of statistical significance varies across models and datasets.

As already noted for random signatures, the center of the distribution and variability of C-indices differ across the datasets. 
Similarly, we investigated whether the prognostic power depends on the number of patients 
and the event rate in a dataset for the reported signatures. As can be seen in the \textit{Supplementary Dataset Dependency}, the MOM and MAD 
are uncorrelated with both the number of patients and with the event rate for reported signatures as well.   	
Thus, these results suggest that the overall prognostic power cannot be increased by collecting a larger number of patients.  

Similar to the results above, Fig. \ref{fig:ridges_signatures} reveals that reported gene-expression signatures 
show an upper C-index limit around 0.8 across all prognostic models and datasets; although the 
highest C-index here is 0.82 (NKI, Ridge).

\section*{Maximum prognostic power of machine learning based gene-expression signatures}

Machine learning has the potential to improve prognostic power, since algorithms such as 
Random Forests have the inherent ability to accommodate interactions between genes \cite{Fernandez-Delgado2014}.
For this reason, we applied 5 state-of-the-art machine learning selection models including
Random Survival Forests with variable importance (SRC), with variable hunting (SRC-VH), 
Minimum Redundancy Maximum Relevance filter (MRMR), and Conditional Variable Importance for Random Forests (CF). 
The univariable model (UM) serves as baseline model, since it selects only one variable
used for prediction of survival (\textit{Material and Methods}). 

The results are given in Fig. \ref{fig:heatmaps} in the form of heatmaps that show the C-indices 
for each combination of machine learning prognostic models (rows) and gene selection methods (columns) for all datasets.

As can be clearly seen, there is not a single prognostic model that outperforms other models.  
Also, there is not a single selection method that clearly outperforms all others.
Both results indicate that there is no gold standard signature that can provide substantially
higher prognostic power, and against which other signatures can be benchmarked.

However, it must be emphasized that we do not aim to find the best machine learning model nor the best selection method, 
but rather to see whether these results approach an universal prognostic limit.

As already noted for random and reported signatures, we investigated whether the prognostic power dependents on the number of patients 
and the event rate in a dataset. \textit{Supplementary Dataset Dependency} shows that the MOM and MAD seem to be uncorrelated 
with both the number of patients and with the event rate, suggesting   	
that the overall prognostic power cannot be increased by collecting more data. 	

Finally, the highest C-index across all datasets is 0.79 (GB-Tree, GSE11121), which repeatedly substantiates 
the existence of a prognostic limit.

\section*{Inherent prognostic limit of gene-expression signatures}

To summarize our results concerning the prognostic power, we plot in 
Fig. \ref{fig:percentage} the percentage of the signatures in all datasets that performs above the C-index indicated on 
the x-axis.

The figure shows that almost all signatures perform above the C-index of 0.3.
Between this value and 0.4 the fraction of signatures starts to steadily decrease,
and a sharp decline is observed in the range 0.4-0.8. The center of this range
is roughly at 0.6 and more than 50\% of signatures perform above this value. 

The number of gene sets exceeding the value of 0.8 is almost 0.
The values around 0.8 deserve special attention and are plotted in detail in the 
inset of Fig. \ref{fig:percentage}. 
From this inspection, we see that the fraction drops below 1\% at the C-index 
of 0.775 for all gene sets, vanishing at 0.825. 
Thus, we estimate the inherent prognostic limit to lie around the C-index $\approx$ 0.8.    
We used 8 datasets, 100 random, 33 reported, and 5 machine learning based signatures,
evaluated by 9 prognostic models, resulting in overall $8\times(100+33+5)\times9 = 9936 \approx 10,000$ signatures. 
In light of these findings, it seems unlikely to find a gene-expression signature that performs
above this limit.

\section*{Missing information in current prognosis}

To quantify and visualize how much information is missing at a specific C-index, we simulated 
survival times based on the MNIST data \cite{Lecun1998} (\textit{Materials and Methods, Evaluation of the missing information}), 
which are 70,000 handwritten $28\times28$ pixel images of digits ranging from 0-9. 
Hereby, we assigned a survival time
to each handwritten digit.
We define the initial amount of information of an image as $100\% - \text{noise} [\%]$.    
In order to reduce the initial amount of information, we added different amounts of noise ranging from 0\% to 100\% to the images and computed 
the C-index at each amount of noise.   
This process was repeated 100 times, from which the median C-index was calculated. 
The results are shown in Fig. \ref{fig:cindex_mnist}.

As can bee seen, prognosis based on images with 100\% noise and with no noise have correctly 
the C-indices of 0.5 and 1.0, respectively. 
Common C-indices reported in the literature range from 0.7 to 0.8 (\textit{Materials and Methods}).
Interestingly, we see that 75\% of the initial information is missing in the middle of this range (C-index=0.75).  
From a more practical perspective, one could argue that a C-index of 0.7 is sufficient for prognosis, 
since one identifies the correct digit from a simple visual inspection. However, more than 60\% of initial 
information is missing at this C-index. Even at the C-index of 0.8 more than 50\% of initial information is missing. 

An alternative but standard way to look at information gain is the Normalized Shannon entropy \cite{shannon1948mathematical}, 
depicted in the inset of Fig. \ref{fig:cindex_mnist}:
$H(x) = - \sum_{k=1}^{N}p_{k}ln(p_{k})/H_{max}$,
where $p_{k}$ is the number of occurrences of the intensity level $k$ divided by the number 
of bins ($28\times28$ pixels for a MNIST image), $N$ denotes the number of intensity levels 
(which is 256 for a gray-scale MNIST image), and $H_{max}$ is the maximum entropy value. 

Thus, $H(x)$ can be interpreted as the amount of randomness in an image $x$. 
For example, $H(x)$ = 0 implies that we know in advance that $p_{k}=1$. Consequently, all pixels will have 
the same value. A value of 0.2 – shown as starting value in the inset of Fig. \ref{fig:cindex_mnist} –
means that we are 20\% uncertain about the information value of the image. 
On the other hand, a Shannon entropy of 1.0 implies that 
we 100\% uncertain about the information content of an image. 

The inset of Fig. \ref{fig:cindex_mnist} shows the dependency of the median C-index 
on the normalized Shannon entropy. Here, following the direction of decreasing entropy, it is 
apparent that the C-index increases drastically from 0.5 to roughly 0.8, meaning that in this range
the C-index can be largely increased by small amounts of information. In contrast, the C-index rises only steadily
above the value of 0.8, hitting a performance plateau, which implies that a prognostic model requires more information gain in order to reach a higher prognostic performance.  

\section*{Discussion}

This study presents two key results. First, we show 
that a maximum prognostic power exists and is inherent to gene-expression signatures. 
We have found a maximum prognostic power as high as 0.8 in terms of the concordance index, meaning that
one can correctly order patients’ prognoses in no more than 80\% of the time.
This result is based on investigating the prognostic power of $\approx$ 10,000 gene-expression signatures developed 
with different methods and tested on different established breast cancer databases.  

While the question of maximum prognostic power has not been previously addressed, these
findings are compatible with what is known about the limits
of predictability in cancer. For example, the review of Lipinski et.al. \cite{Lipinski2016} 
outlines the role of stochastic factors in cancer evolution
that fundamentally limit the predictability and development of more accurate prediction algorithms.
Other works \cite{Stegemann2007, Burrell2013, Blount2018} also recognized a theoretical limit
on predictability in cancer evolution. However, not a single study has provided an empirical
estimation so far, which is of practical relevance in clinical prognosis.  

\begin{figure*}[tbhp]
	\centering
	\includegraphics[width=1.0\textwidth]{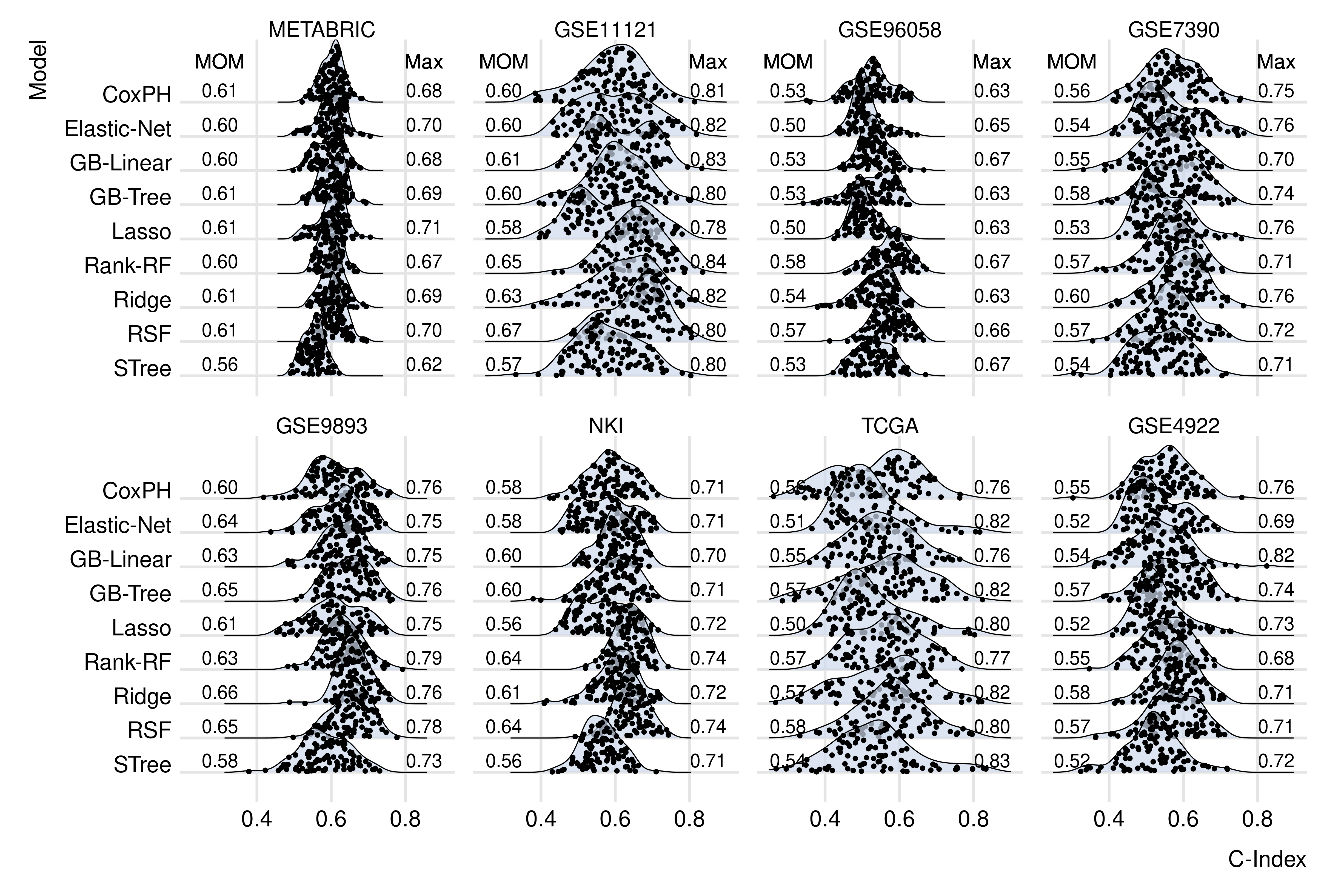}
	\caption{\textbf{Random signatures}: Distributions of C-indices computed based on random gene sets. There are
		overall 100 signatures evaluated for each prognostic model (row).   
		A single dot denotes the median C-index resulting from a prognostic model based on a single 
		random selection and computed by the 5-fold cross-validation. Each gene set contains expression values of 20 genes.
		MOM represents the median of sample medians and MAX is the corresponding maximum C-index. 
		Each row represents the survival model used for the computation (\textit{Materials and Methods}): 
		Cox proportional hazards model (CoxPH),
		Lasso regression (Lasso), Ridge regression (Ridge), elastic net survival regression (Elastic-Net),
		Gradient boosting with linear learners (GB-Linear), with tree-based learners (GB-Tree),
		Random survival forests (RSF), maximally selected rank statistics random forests (Rank-RF), and survival trees (STree).}
	\label{fig:ridges_random}
\end{figure*}

\begin{figure*}[tbhp]
	\centering
	\includegraphics[width=1.0\textwidth]{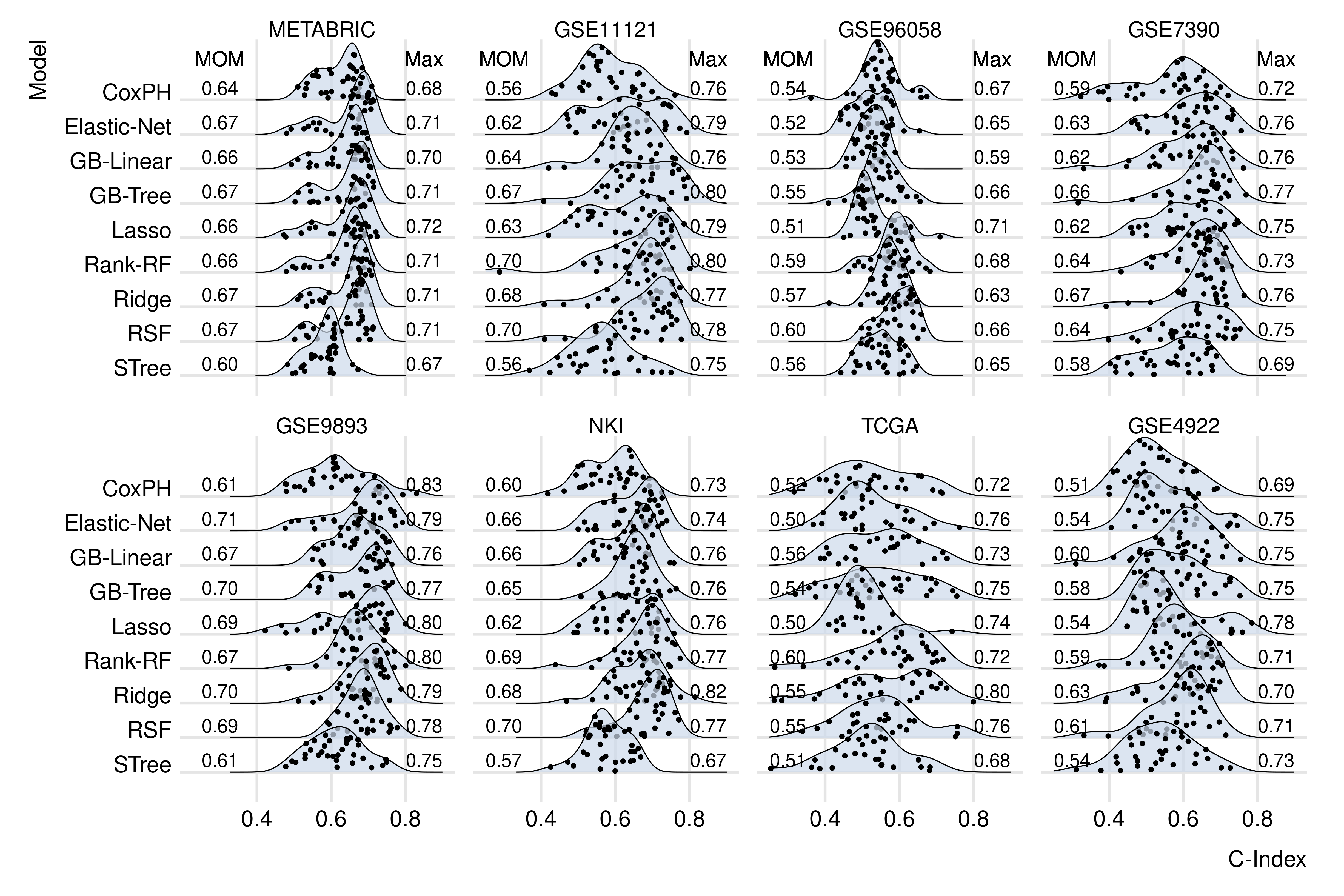}
	\caption{\textbf{Reported signatures}: Distributions of C-indices computed based on reported gene sets. 
		There are overall 33 signatures for each prognostic model (row).
		A single dot denotes the C-index resulting from a model based on a single 
		gene set. Gene sets contain various number of genes.
		Each row represents the survival model used for the computation (\textit{Materials and Methods}): 
		Cox proportional hazards model (CoxPH),
		Lasso regression (Lasso), Ridge regression (Ridge), elastic net survival regression (Elastic-Net),
		Gradient boosting with linear learners (GB-Linear), with tree-based learners (GB-Tree),
		Random survival forests (RSF), maximally selected rank statistics random forests (Rank-RF), and survival trees (STree).}
	\label{fig:ridges_signatures}
\end{figure*}

\begin{figure*}[tbhp]
	\centering
	\includegraphics[width=1.0\textwidth]{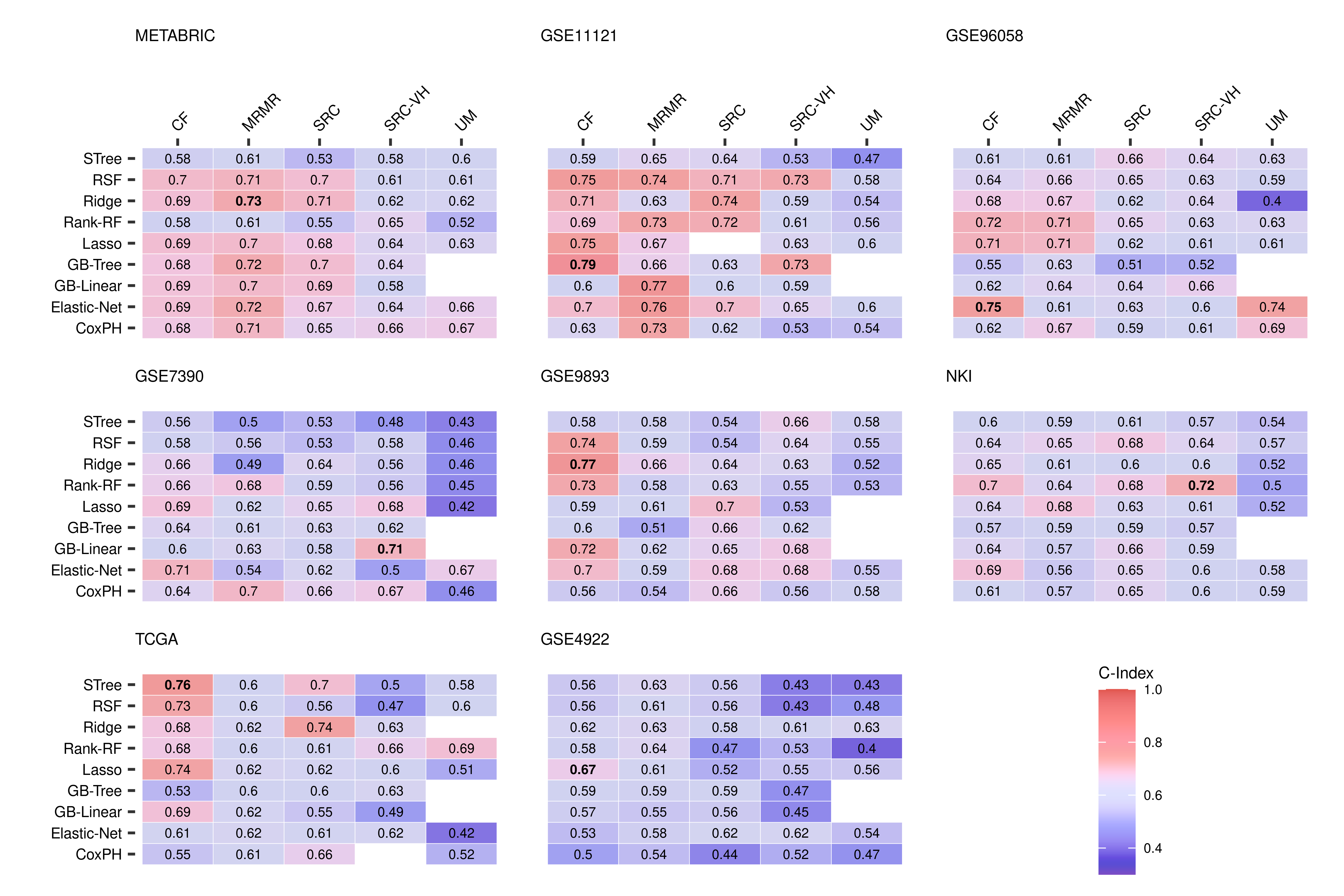}
	\caption{\textbf{C-indices of gene sets selected with 
			machine learning.} Gene sets contain various number of genes (\textit{Materials and Methods}).
		Each row represents prognostic model used for the computation: Cox proportional hazards model (CoxPH),
		Lasso regression (Lasso), Ridge regression (Ridge), elastic net survival regression (Elastic-Net),
		Gradient boosting with linear learners (GB-Linear), with tree-based learners (GB-Tree),
		Random survival forests (RSF), maximally selected rank statistics random forests (Rank-RF)
		, and survival trees (STree). Each column represents selection method used for the computation:
		conditional variable importance for random forests (CF), random survival forests with variable importance
		(SRC), with variable hunting (SRC-VH), minimum redundancy maximum relevance filter (MRMR),
		and the univariate model (UM).}
	\label{fig:heatmaps}
\end{figure*}

\begin{figure}[tbhp]
	\centering
	\includegraphics[width=1.0\linewidth]{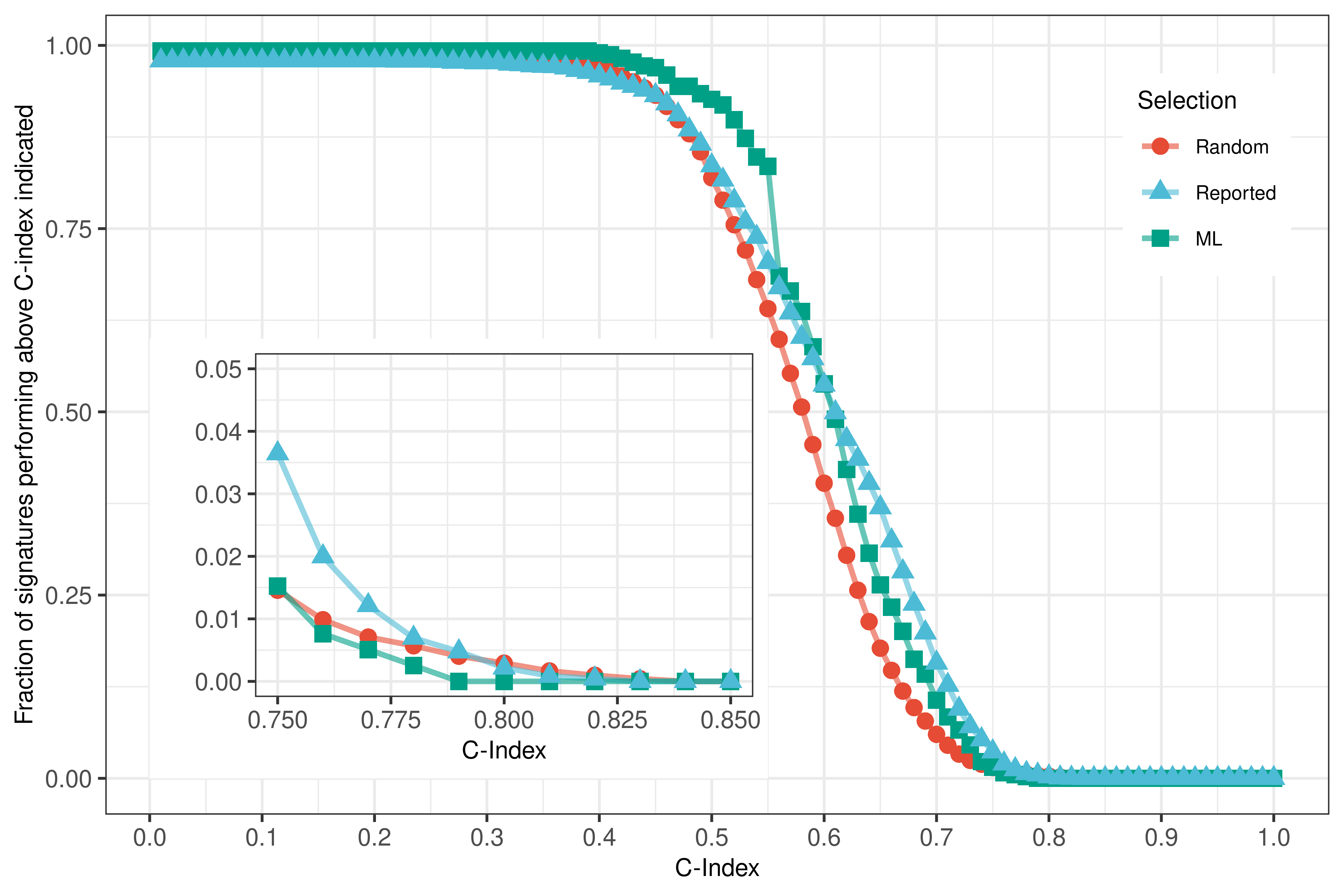}
	\caption{\textbf{Fraction of gene sets performing above C-index indicated on the x-axis.} For 
		each selection approach – i.e. for random, reported, and machine learning 
		based (ML) selections  –  gene sets from all 8 datasets were used.}
	\label{fig:percentage}
\end{figure}

There are also differences in C-indices between datasets, as observed in Fig.s \ref{fig:ridges_random}
and \ref{fig:ridges_signatures}, which may be related to the inter-platform and inter-cohort variability.
However, a more probable explanation may be the difference in event types used for prognosis. More specifically,
the prediction of disease-free, distant-metastasis-free, and recurrence-free survival is more specific than
the prediction of the overall survival, which may include events that are not related to the disease. 
The \textit{Supplementary Event Type} provides conclusive support for this explanation, where box plots
of median C-indices are shown across datasets, indicating that both datasets – where the overall survival
is used for the prognosis – exhibit the lowest performances. 

One may assume that the prognostic power of gene-expression signatures would 
increase by collecting a larger number of patients. 
However, our results demonstrate that the maximum prognostic power is independent on the number 
of patients. One could also argue that the sample sizes – the largest dataset contains $\approx700$ patients – 
are not sufficient to proof this. 
For this reason, we combined the 8 datasets into one large dataset resulting in 2500 patients, 
resampled the data with different sample sizes ranging from 800 to 2500 patients, 
and measured the median of the sample medians of 1000 random signatures at each sample size (\textit{Supplementary Combined Dataset}). 
As can be seen there,
the performance does not increase with larger sample sizes ($p = 0.33$).           
Moreover, the authors of \cite{Yousefi2016} 
provide evidence that prognostic power cannot be increased with more data, although they measured 
the classification error of prognostic models instead of the concordance index. 
This points toward that not random noise hinders a better prediction, but that not all information 
is available.

\begin{figure*}[tbhp]
	\centering
	\includegraphics[width=1.0\textwidth]{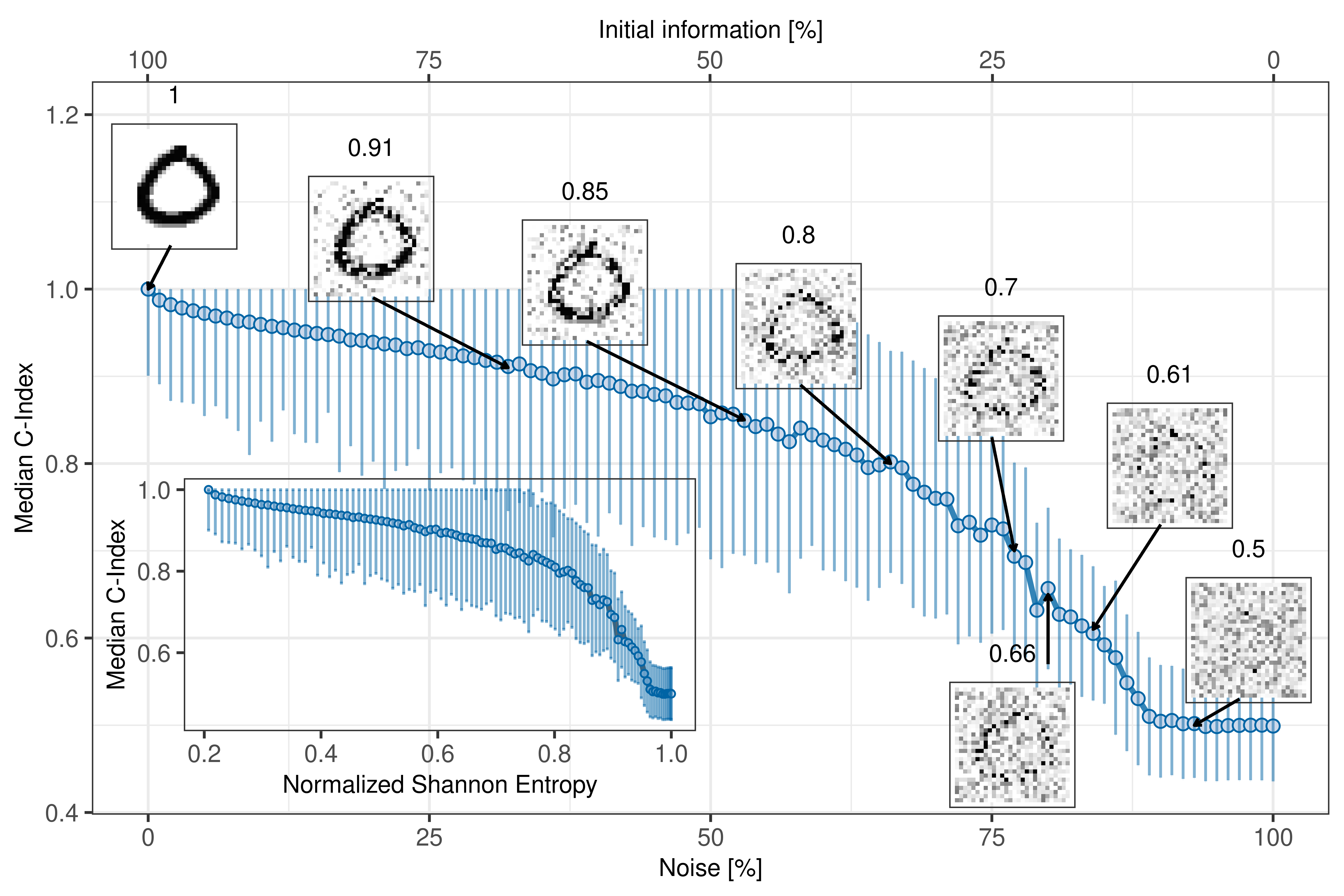}
	\caption{\textbf{Median C-indices of simulated survival time predictions based on MNIST data.}
		We sample 2000 MNIST images at random, add different amount of noise to images, 
		then reduce the images to 2-dimensional data points, which are used in the Cox proportinal hazards models 
		to simulate survival times.
		Finally, we compute the C-indices. This procedure is repeated 100 times and statistics are recorded.
		The simulated survival times of images without noise are used as true survival times.
		Initial information of an image is $100\% - \text{noise} [\%]$.
		The number above exemplary images indicate calculated C-index of the corresponding image. 
		In the inset, dependency on the normalized Shannon entropy is depicted. Error bars indicate
		the standard error of the median based on 100 random samples.	
	}
	\label{fig:cindex_mnist}
\end{figure*}

The second key result of this study is the amount of information that is missing in gene signatures 
to provide a prognosis with a concordance above 80\%.
Quantitatively, it can be inferred from our interpretation of the concordance index in Fig. \ref{fig:cindex_mnist},
that at a maximal C-index of 0.8, there are still 50\% of missing information.
This result suggests that we are currently still quite far from harnessing all information required for an
optimal prognosis.
 
If a significant lack of prognostic information exists, what factors have the potential to improve prognosis?  

As can be seen on our results, models based on a single
variable (univariable models) show the lowest prognostic power. 
Thus, it is unlikely that a single gene-expression, clinical or histological variable will capture the missing information and determine prognosis with high prognostic power. 
The continued refinement of prognostic models that incorporate many complementary factors may
allow for more accurate predictions of outcome. 	
For instance, we have recently shown that the prognostic power can be substantially improved by simply combining
gene expressions and clinical information into a hybrid model \cite{Tschodu2022}, although the same prognostic 
limitation may apply to other variables such as clinical and histological parameters. 
Similarly, multiple studies in cancer prognosis 
have recently emphasized a hybrid approach as well \cite{Popovici2016, Gallins2020, Hao2020, Schneider2022}.

Moreover, studies on gene-expression signatures have focused on defining molecular determinants within the tumor environment \cite{karn2015influence}.
However, host factors such as immune response, dietary variables, or hormone milieu may have profound effects 
on cell proliferation, invasion, and metastasis.   
Consequently, models that combine so-called tumor factors and host factors 
could allow to harness the missing information and substantially enhance the prognostic power.

However, combining complementary factors may not be sufficient and raises the question as to whether complementary combinations can capture
the potential missing information. There are more complex relations, which cannot be clarified by a simple summation
of orthogonal factors. We used machine learning models that can detect complex, non-linear effects between genes. However,
we cannot be certain that these effects were indeed detected. 
In this context, such complementary factors may be other emergent effects that are not very sensitive to gene expression changes.
For example, the migration of cancer cells through dense tissues is strongly dependent on the biomechanical interactions of 
the cancer cells with the host tissue \cite{grosser2021cell}, which at best only indirectly by mechanobiology depend on the cancer cells' gene expression. We have recently introduced the so-called cancer cell unjamming as a simple biomechanical
cell motility marker that describes the squeezing of migrating cells in dense
tissue (\url{https://doi.org/10.21203/rs.3.rs-1435523/v1}). The unjamming  within primary tumors as an emergent physical phenomenon is part of the metastatic cascade, and is complementary to the current prognosis, since no other marker reliably accounts for cell motility \cite{gerashchenko2019markers}.
From a clinical perspective there is an urgent need to precisely stratify patients with special attention on the question of whether any adjuvant measure after surgery is indicated. In current clinical practice, this is done on the basis of the histological tumor type, grade, and stage, and in addition under consideration of the results of an immunohistochemical workup as well as the determination of molecular biomarkers including gene expression profiles. Our approach could, after careful clinical validation, have an immediate impact by adding a new dimension, i.e. a morphological surrogate marker for motility, to the assessment of the prognosis of an individual breast cancer patient. And since this all would be done on a standard H\&E stained slide, except for the software no further investment would be necessary.

Several commercially available gene-expression signatures are frequently used in the clinic not only for prognosis, but
also to guide the decision whether a patient would benefit from adjuvant chemotherapy, i.e. to decide whether one can
spare a potentially toxic treatment to a patient for whom it is not likely to be beneficial \cite{varnier2021using}. 
Our results demonstrate that a prognosis based on gene-expression
signatures alone is lacking 50\% of the potential information and is thus fundamentally limited. However, it can be substantially improved by other relevant patient information, since more than the half of information is left to be revealed.

\newpage

\bibliography{main.bib}

\bibliographystyle{Science}

\section*{Acknowledgments}

We thank Bernhard Ulm and Dr. med. Benjamin Wolf for valuable feedback. 
\textbf{Author contributions:}
Conceptualization: D.T., J.L., P.G., A.N., and J.A.K.
Methodology: D.T., J.L., P.G.,
Software: D.T.
Formal analysis: D.T.
Data Curation: D.T.
Writing -- original draft preparation: D.T.
Writing --review and editing: D.T., J.L., A.S.W., A.N., and J.A.K.
Supervision: A.N., and J.A.K.
Project administration: A.N., and J.A.K.
\textbf{Competing interests:}
The authors declare no competing interests.
\textbf{Data and Materials availability:}
All datasets – except METABRIC – are available in the NCBI Gene Expression Omnibus: \url{https://www.ncbi.nlm.nih.gov/geo/}.
METABRIC data are available from the METABRIC Institutional Data Access/Ethics Committee (contact via metabric[at]cruk[dot]cam[dot]ac[dot]uk) for researchers who meet the criteria for access to confidential data. The data
underlying the results presented in this study are available from \url{https://ega-archive.org/dacs/EGAC00001000484}.
Code scripts to download, prepare, and analyze data are deposited in a Github repository: \url{https://github.com/DiTscho/LimitOfPrognosis}.   
\textbf{Funding:}
This work was supported by the European Research Council (ERC
741350/HoldCancerBack, \url{https://cordis.europa.eu/project/id/741350/de}), the DeutscheForschungsgemeinschaft (DFG KA 1116/17-1 and
INST 268/296-1 FUGG, \url{https://www.dfg.de/}), and the National Science Foundation (NSF-PHY1607416, \url{https://www.nsf.gov/awardsearch/showAward?AWD_ID=1607416}).

\section*{Supplementary Materials:}

Materials and Methods \\
Supplementary Data \\
Supplementary Reported Gene Selections \\
Supplementary Machine Learning Models \\
Supplementary Gene Selection Methods \\
Supplementary Random Signature Size \\
Supplementary Random Signature Superiority \\
Supplementary Difference Between Selection Methods \\
Supplementary Dataset Dependency \\
Supplementary Combined Dataset \\
Supplementary Event Type \\
Figs. S1 to S8 \\
Tables S1 to S5

\end{document}


	
	\baselineskip24pt
	
	
	\maketitle

	\section*{This PDF file includes:}
	
	\begin{spacing}{1.125}
	Materials and Methods \\
	Supplementary Data \\
	Supplementary Reported Gene Selections \\
	Supplementary Machine Learning Models \\
	Supplementary Gene Selection Methods \\
	Supplementary Random Signature Size \\
	Supplementary Random Signature Superiority \\
	Supplementary Difference Between Selection Methods \\
	Supplementary Dataset Dependency \\
	Supplementary Combined Dataset \\
	Supplementary Event Type \\
	Figs. S1 to S8 \\
	Tables S1 to S5
	\end{spacing}

	\section*{Materials and methods}
	
	\subsubsection*{Datasets}\label{datasets}	
	
	The stability of a variable selection method can be affected by
	changes in the data and varies between datasets, i.e. a selection method can produce gene sets that will be 
	different or invalid when changes to the data occur or a new dataset is used \cite{Dittman2013}.  
	Thus, gene expressions along with survival data are collected form 8 different datasets.
	
	A dataset can be described as a matrix whose columns contain expression values of thousands of genes
	and whose rows are organized by tissue samples.
	The expression values are produced by a quantification method such as reverse transcriptase-PCR, as used for quantification
	in the Oncotype DX and EndoPredict signatures, or DNA-microarray technology, as used for quantification in the MammaPrint signature.  
	Previous studies have shown that the stability of gene selections varies across 
	different datasets \cite{Dittman2013, Haury2011}. 
	Therefore, we use 8 different but well-established datasets:
	METABRIC \cite{Curtis2012a}, GSE9893 \cite{Chanrion2008a}, GSE7390 \cite{Desmedt2007a}, 
	GSE96058 \cite{Brueffer2018b}, GSE11121  \cite{Schmidt2008a}, GSE4922 \cite{Ivshina2006a}, and data generated
	by the TCGA Research Network: \url{https://www.cancer.gov/tcga}.

	The signatures reported in the literature have been proven to work well in hormone receptor-positive breast
	cancers \cite{Huang2018}.
	Thus, estrogen positive patients were selected who did not receive chemotherapy to avoid therapy effects as well.  
	Patients above the age of 70 years are disregarded to avoid events due to concomitant diseases.  
	
	A detailed description of data processing and descriptive statistics of each dataset including the number of events and the event used for a specific dataset are provided 
	in \textit{Supplementary Data}.

	\subsubsection*{Prognostic models}\label{survival_analysis}
	
	Survival analysis is a statistical treatment of data that aims to predict the time from a fixed time point 
	to an event such as the time from surgery to death. The inherent aspect of survival analysis is the presence 
	of censored data, indicating that the event of interest is never observed in all patients.   
	For example, a patient may be lost to followup or the event does not occur within the study duration.
	The most common method for analyzing these censored data has been the Cox proportional hazards model (CoxPH). 
	
	CoxPH model the relationship between the outcome and several variables (also called covariates) by computing the
	following hazard function:

	\begin{equation*}
		h(t) = h_{0} \times \exp(\beta_{1}x_{1} + \beta_{2}x_{2} + ... + \beta_{n}x_{n}),
	\end{equation*}
	
	which represents the hazard of an event occurring at time point $t$ by assigning a (risk) coefficient $\beta_{i}$ to each variable $x_{i}$,
	whereas $h_{0}(t)$ is the baseline function that is unspecified since it vanishes by dividing the hazards of different patients. 
	By adding the hazards to a time point, the cumulative hazard can be computed and is used to estimate the probability
	of an event occurring.    
	As can be seen in the equation above, the Cox model assumes that the effects of covariates on survival are additive and constant over time.

	However, the Cox model does not generalize well to high dimensional data, where the number of variables exceeds
	the number of patients. For this reason, machine learning models that extend CoxPH can be applied.
	The Lasso model (Lasso), Ridge model (Ridge), and the Elastic net model (Elastic-Net) are extensions of CoxPH that
	incorporate so-called penalties. These penalties are often used to shrink the risk coefficients 
	so that less important variables have less effect in the model.  
	With LASSO penalty a model with a smaller set of coefficients is produced, whereas with Ridge penalty all 
	coefficients are shrinked by the same factor. The Elastic-Net is a linear combination of both penalties
	and is used to overcome several limitations of Lasso and Ridge (see \textit{Supplementary Machine Learning Models}).

	

	CoxPH and its penalized extensions impose strong assumptions
	on the hazard function, that variables are additive and relate multiplicatively to the hazard, as
	well as that hazard remains constant over time. However, these assumptions are often violated in high-dimensional time-to-event data. 
	To alleviate these problems, algorithms based on random forests and gradient boosting machines can be used.  
	CoxPH, Lasso, Ridge, and  Elastic-Net are single predictive models. Random forests and gradient
	boosting machines, on the other hand, are ensemble learning algorithms that combine multiple predictive models into an overall ensemble. 
	
	For this, original data are resampled by drawing samples with replacement. This procedure is known as bootstrapping.  
	Then a prediction model – called base learner – is applied to each bootstrap sample, and predictions are
	made by averaging the predictions from the individual base learners. This method is referred to as bagging.
	
	Random forests utilize bagging of decision trees as base learners. Decision trees is an algorithm that 
	recursively applies a set of yes/no rules to split variables and make predictions based on these splits. 
	However, bagging of decision trees results in tree correlation since samples are drawn with replacement, 
	and thus are partially redundant.
	Random forests decorrelates decision trees by performing each split on a random subset of the original 
	variables.

	Survival trees (STree) and random survival forests (RSF) are extensions of decision trees and random forests 
	to censored time-to-event data, respectively.
	More specifically, RSF maximize the survival difference to find the best split of variables. This 
	is done by maximizing the log-rank statistic over all available split points and variables.
	A major drawback of random forests and RSF is the bias toward selecting variables with many possible 
	split points to splits on, e.g. a variable with larger variance.  
	Maximally selected rank statistics random survival forests (Rank-RF) overcome this drawback
	by separating the selection of the variable to split on from the selection of the split point. Instead, a split
	point is chosen  using maximally selected rank statistics, which can also identify non-linear effects in variables. 
	
	Whereas random forests combines independent base learners, gradient boosting machines combines simple base learners,
	e.g. a decision tree with a few splits, that are trained sequentially in order to improve (boost) upon the performance
	of the predecessor.  
	Gradient boosting is trained on the residual errors (gradients) of the entire ensemble at each learning step. 
	Gradient Boosting with linear boosting (GB-Linear) is trained with linear models such as CoxPH as base learners,
	whereas tree-based boosting (GB-Tree) is trained with decision trees as base learners.


	All models and model parameters are described in \textit{Supplementary Machine Learning Models}. 
	

	\subsubsection*{Selecting gene sets with machine learning}\label{selecting_gene_sets}
	
	Variable selection – also called feature selection – is frequently used as a preprocessing to machine learning. 
	It is a process of choosing a subset of original variables in order to remove irrelevant and redundant variables, 
	and thus improve learning performance.  
	In recent years, however, especially gene expression data have become increasingly larger in both number of 
	patients and number of variables containing high degree of irrelevant and redundant information that may greatly degrade
	the performance of learning algorithms \cite{Dittman2013}. Therefore, variable selection is necessary
	for handling high-dimensional data.
	
	We use one standard and four different machine learning gene selection methods.
	
	Univariable model (UM) uses the univariable CoxPH model that includes just one variable, namely expression values of a single gene, 
	to model the outcome, i.e. the survival of a patient.  After each gene in a dataset is modeled to 
	the outcome, gene with the best prognostic performance is chosen.
	
	Random survival forests with variable importance (SRC), 
	with variable hunting (SRC-VH), minimum redundancy maximum relevance filter (MRMR), 
	and conditional variable importance for random forests (CF) use random survival forests to 
	model the outcome but incorporate different measures, i.e. different splitting criteria, for variable importance. 
	
	The variable importance of the random forests algorithm (SRC) is computed by permuting the expression
	values of each gene and calculating the difference between the performances of the prognostic model before and after permutation. 
	Subsequently, the genes are ranked based on these differences and a specified threshold is used to select the most important genes.   
	
	The survival random forests variable hunting method (SRC-VH), on the other hand, use a different importance score.
	First, the standard variable importance, i.e. SRC, is performed.  
	Second, a random subset of genes is selected with probability proportional to calculated variable importances, and a 
	prognostic model is build.  
	Third, the selected genes are ordered by the shortest distance from the tree root to the largest subtree including this gene as its root; they are added successively to the prognostic model until the joint importance does not increase anymore.
	These steps are iterated a specified number of times. Eventually, the variable importances result from the ranking
	of the variables based on the frequency of occurrence in these iterations. 
	
	The minimum redundancy maximum relevance algorithm (MRMR) \cite{radovic2017minimum} selects variables that are mutually far away from each other,
	since variables that are mutually close to each other might be redundant. Thus, the algorithm minimizes redundancy by removing the potentially redundant variables. At the same time, the selected variables are highly correlated with the outcome, meaning that they exhibit maximum relevance.
	
	As already mentioned, a major drawback of random forests and RSF is the bias toward selecting variables with many possible. 
	To address this problem, conditional variable importance for random forests (CF) that utilizes the linear rank statistic as splitting criterion
	can be used as well.  
	
	All models and parameters used are described in detail in \textit{Supplementary Gene Selection Models}.

	\subsubsection*{Concordance index}\label{cindex}

	To assess the prognostic power, we choose the C-index – also called the concordance index or Harrel's 
	C-index \cite{Harrell1982, Harrell1984} – since it is a standard measure for evaluating survival times \cite{Penciana2004}
	and  prognostic groups with short-term or long-term survivors can be confidently 
	constructed from the C-index.
	For instance, it was used by the 2012 DREAM Breast Cancer Challenge that was designed to improve survival 
	prediction (\url{https://doi.org/10.7303/syn1710250}, \cite{Margolin2013}). 
	In this challenge, clinical and genomic data of around 2000 patients were available. 
	
	The C-index describes the ability of a prognostic model
	to separate patients with good and poor outcomes. 
	It is a common practice to recall that a C-index of 0.5 denotes a completely random prognosis 
	and a value of 1.0 implies that one can perfectly discriminate 
	predicted patients survival probabilities according to their survival times: a patient with a 
	higher survival time would get a higher probability than a patient with a shorter survival time. 
	A C-index = 0 describes the perfect anti-concordance, where the predicted survival probabilities 
	are inversely proportional to survival times.  
	
	In his seminal work Harrell \cite{Harrell1982} provides the interpretation of the C-index as percentage of patients 
	that can be correctly ordered. For instance, a value of 0.7 indicates that one can correctly order patients' prognoses 
	70\% of the time. Starting from this work, surprisingly, the interpretation of the intermediate C-indices from 0.5 to 1.0 
	has not been considered more closely \cite{Longato2020}. More recently Longato et.al. \cite{Longato2020} addressed this 
	problem and proposed a simplified view on the C-index by relating its values to the number of patients whose scores are in the correct order relative to their survival times. 
	Below we provide a further simplified view on the C-index that establishes a relationship between its values and the content of missing information. 
	
	\subsubsection*{Evaluation of the missing information}
	
	To visualize and quantify how much information is missing at a specific C-index, 
	we adapt an idea from \cite{Gensheimer2019} and simulate a scenario, where a 2D image 
	of a handwritten digit corresponds to one patient. We used the MNIST dataset 
	\cite{Lecun1998} consisting of 70,000 handwritten $28\times28$ pixel images of digits ranging from 0-9. 
	Then we successively remove information by adding noise to these images, as can be exemplary seen in the images 
	from left to right above the graph in Fig. \ref{fig:cindex_mnist}.

	Authors of \cite{Gensheimer2019} simulated survival times based on integer digits ranging from 1 to 4, 
	so that patients with higher digits tend to have shorter survival times. Consequently, their images 
	may represent X-ray images of tumors, with higher digits representing larger, 
	more deadly tumors. 
	In contrast, we choose continuous values in order to model gene-expressions.   
	
	First, we randomly sample 2000 images, since this number corresponds roughly to the number of 
	patients in a large enough BC gene expression dataset such as METABRIC and the dataset from 
	the DREAM Breast Cancer Challenge. 
	
	Second, we reduce these 784-dimensional ($28\times28 = 784$) images  to
	2-dimensional continuous data points that can be imagined as expression values
	of 2 different genes; and simulate survival times. The reduction is done by applying 
	the Principal Component Analysis, which is a technique of reducing high-dimensional data 
	to new uncorrelated variables – called principal components (PCs) – by maximizing variance, 
	i.e. minimizing information loss.    
	
	We simulate the survival time $T(x)$ for each image $x$ with the following expression:
	$T(x) = \text{MST} \cdot \exp (\text{S} \cdot \text{(PC1 + PC2)})$, 
	where PC1 and PC2 denote the values of the first and second PCs, respectively.
	Here, a higher value pf PC1 or PC2 could be imagined as a higher expressed gene.
	
	MST represents the median survival time and S is the coefficient that regulates the skewness of 
	the simulated survival time distribution: the larger this parameter the more right-skewed 
	is the distribution. 
	
	To simulate a more realistic survival scenario,  we set MST and S to 10 years and to 0.6, respectively,
	since the value of 0.6 corresponds to most real survival time distributions that are right-skewed. 
	
\begin{figure*}[htp!]
		\centering
		\includegraphics[width=1.0\linewidth]{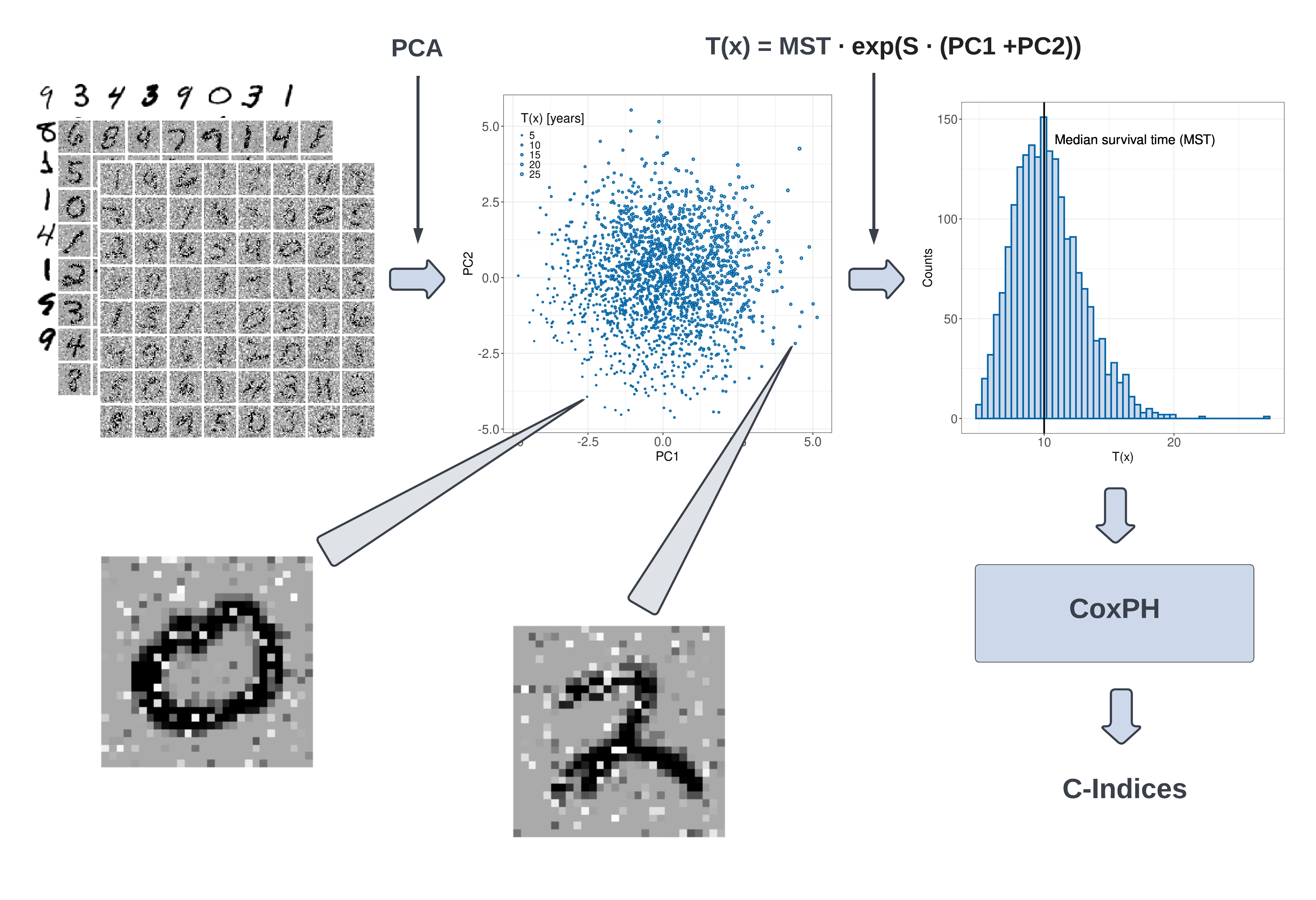}
		\caption{\textbf{Simulation of survival times.} (top left) We sample 2000 MNIST images at random and add different amounts of
			noise ranging from 0\%-100\%. (top middle) For each amount of noise, the sample is reduced to 2 dimensions (PC1 and PC2) using the
			principle component analysis (PCS). (bottom left and bottom right) Exemplary noised images: each image is reduced
			to a single point with PC1 and PC2 as coordinates. (top right) Exemplary distribution of survival times based on PC1 and PC2 after 
			applying the equation shown above the distribution. (bottom right). Principal components
			serve as input variables to the Cox proportional hazards model and simulated survival times as outcome. Eventually,
			C-index is computed at each amount of noise.}
		\label{fig:missing_info}
	\end{figure*}

	In the last step, these reduced data are used as variables in the Cox model, from which 
	the C-index is eventually calculated. This procedure is depicted in Fig. \ref{fig:missing_info}.
	
    \section*{Supplementary Data}

    For downloading or preprocessing data, Bioconductor packages (\url{https://www.bioconductor.org/}) in R (\url{https://www.r-project.org/}) 
    were used. 
    
    All datasets that are available in the NCBI Gene Expression Omnibus \cite{barrett2005ncbi} – i.e. all datasets except METABRIC, TCGA, and NKI – were retrieved using the R package GEOquery \cite{davis2007geoquery}.
    If available and if not otherwise mentioned, we downloaded the raw .CEL files, standardized them with the RMA procedure \cite{irizarry2003exploration}, and filtered the data using the WGCNA procedure as described in \cite{langfelder2008wgcna}. 
    
    We used ComBat in the R package \textit{sva} \cite{leek2012sva} to adjust data for batch effects.
    
    All scripts to download and prepare data are deposited in a Github repository: \url{https://github.com/DiTscho/LimitOfPrognosis}.    
    
    In all datasets, estrogen-receptor positive (ER+) patients under the age 70 who did not receive cytotoxic chemotherapy were 
    selected. 
    
    In all datasets, genes without annotations were removed.

    \subsection*{Datasets}

    Table \ref{table:descriptive_stats} summarizes common important clinical parameters across all datasets. 
    
    	\begin{table}[htp!]
   
    	\caption{Descriptive statistics of common important clinical parameters for all datasets used in this study.}
    	\label{table:descriptive_stats}
    	\centering
    	\scalebox{0.7}{
    	\begin{tabular}{lllll}
    		\hline
    		& GSE11121 (N=120) & GSE7390 (N=134) & GSE96058 (N=147) & METABRIC (N=683) \\ 
    		\hline
    		age &  &  &  &  \\ 
    		-  Median & 58.000 & 47.000 & 64.000 & 58.140 \\ 
    		-  Q1,Q3 & 48.000, 63.250 & 42.250, 51.000 & 56.500, 67.000 & 50.255, 63.845 \\ 
    		grade &  &  &  &  \\ 
    		-  1 & 18 (15.0\%) & 29 (22.0\%) & 39 (26.5\%) & 89 (13.5\%) \\ 
    		-  2 & 86 (71.7\%) & 68 (51.5\%) & 81 (55.1\%) & 329 (49.9\%) \\ 
    		-  3 & 16 (13.3\%) & 35 (26.5\%) & 27 (18.4\%) & 241 (36.6\%) \\ 
    		tumor size &  &  &  &  \\ 
    		-  Median & 1.800 & 2.000 & 1.600 & 2.000 \\ 
    		-  Q1,Q3 & 1.400, 2.325 & 1.600, 2.500 & 1.200, 2.200 & 1.560, 2.700 \\ 
    		node status &  &  &  &  \\ 
    		-  0 & 120 (100.0\%) & 134 (100.0\%) & 101 (68.7\%) & 433 (65.0\%) \\ 
    		-  1 & 0 (0.0\%) & 0 (0.0\%) & 40 (27.2\%) & 233 (35.0\%) \\ 
    		-  NA & 0 (0.0\%) & 0 (0.0\%) & 6 (4.1\%) & 0 (0.0\%) \\ 
    		survival time in years &  &  &  &  \\ 
    		-  Median & 7.583 & 12.731 & 4.071 & 11.633 \\ 
    		-  Q1,Q3 & 5.500, 10.729 & 7.764, 15.504 & 3.110, 5.263 & 7.001, 17.683 \\ 
    		event &  &  &  &  \\ 
    		-  0 & 97 (80.8\%) & 108 (80.6\%) & 94 (63.9\%) & 519 (76.0\%) \\ 
    		-  1 & 23 (19.2\%) & 26 (19.4\%) & 53 (36.1\%) & 164 (24.0\%) \\ 
    		
    		\midrule
    		
    		& GSE4922 (N=80) & GSE9893 (N=86) & NKI (N=179) & TCGA (N=169) \\ 
    		\hline
    		age &  &  &  &  \\ 
    		-  Median & 57.000 & 60.550 & 46.000 & 61.038 \\ 
    		-  Q1,Q3 & 48.750, 66.250 & 55.325, 65.650 & 41.000, 50.000 & 51.712, 64.882 \\ 
    		grade &  &  &  &  \\ 
    		-  1 & 33 (41.2\%) & 15 (18.1\%) & 52 (29.1\%) & 0 \\ 
    		-  2 & 39 (48.8\%) & 55 (66.3\%) & 67 (37.4\%) & 0 \\ 
    		-  3 & 8 (10.0\%) & 13 (15.7\%) & 60 (33.5\%) & 0 \\ 
    		tumor size &  &  &  &  \\ 
    		-  Median & 1.700 & 2.000 & 2.000 & NA \\ 
    		-  Q1,Q3 & 1.200, 2.200 & 1.600, 2.500 & 1.500, 2.500 & NA \\ 
    		node status &  &  &  &  \\ 
    		-  0 & 79 (98.8\%) & 46 (54.8\%) & 131 (73.2\%) & 109 (64.9\%) \\ 
    		-  1 & 1 (1.2\%) & 38 (45.2\%) & 48 (26.8\%) & 59 (35.1\%) \\ 
    		survival time in years &  &  &  &  \\ 
    		-  Median & 10.292 & 5.496 & 6.521 & 0.197 \\ 
    		-  Q1,Q3 & 5.688, 10.771 & 4.681, 6.567 & 4.711, 9.749 & 0.132, 0.366 \\ 
    		event &  &  &  &  \\ 
    		-  0 & 54 (67.5\%) & 57 (66.3\%) & 123 (70.3\%) & 152 (89.9\%) \\ 
    		-  1 & 26 (32.5\%) & 29 (33.7\%) & 52 (29.7\%) & 17 (10.1\%) \\ 
    		\hline
    	\end{tabular}
    }
    \end{table}

    \subsubsection*{METABRIC}
    
    Clinical and pathological annotations as as well as gene expression of over 2000 breast cancer tumors were obtained by 
    permission from the METABRIC (Molecular Taxonomy of Breast Cancer International Consortium) consortium \cite{Curtis2012a}.
    These data can be downloaded from the EuropeanGenome-Phenome Archive at \url{http://www.ebi.ac.uk/ega} under accession number EGAS00000000083. The gene expression values were measured on the Illumina HT-12 v3 platform, already preprocessed and log2-normalized, as reported in \cite{Curtis2012a}. The function \textit{avereps} in the R package \textit{limma} \cite{Ritchie2015} was used to summarize genes with multiple probes. The R package \textit{illuminaHumanv3.db} was used to annotate genes \cite{dunning2015illuminahumanv4}. 
    From the initial 2136 samples we selected 683 samples of patients, who either died due to the disease or are still alive.

    \subsubsection*{TCGA (The Cancer Genome Atlas)}
    Breast cancer RNA Seq gene expression and clinical data were downloaded  
    from the TCGA website (\url{http://cancergenome.nih.gov}) using the package TCGAbiolinks \cite{colaprico2016tcgabiolinks}. 
    Gene expression were filtered  and normalized w.r.t. normal solid tissue by the \textit{TMM} method from the \textit{edgeR} R package 
    \cite{robinson2010edger} and the \textit{voom} method \cite{law2014voom}. From the initial 1095 samples, we selected 169 using
    the criteria mentioned above. The overall survival time and status were used for survival analysis.

    \subsubsection*{GSE11121}
    The datasets contains 200 samples of consecutive lymph node-negative breast cancer patients treated at the Department of Obstetrics and Gynecology of the Johannes Gutenberg University Mainz between 1988 and 1998 \cite{Schmidt2008a}.
    From the initial 200 samples, we selected 120 using
    the criteria mentioned above. The distant-metastasis-free survival time and status were used for survival analysis.

    \subsubsection*{GSE96058}
    
    The available expression matrix contained preprocessed log2-normalized expression values of a prospective population-based series 
    of 3,273 BC patients with a median follow-up of 52 months (Sweden Cancerome Analysis Net- work—Breast [SCAN-B], ClinicalTrials.gov identifier: NCT02306096), as described in \cite{Brueffer2018b}. No further standardization was conducted.
    From the initial 3,273 samples, we selected 1102 using
    the criteria mentioned above. We subsequently performed downsampling due to a low amount of events ($< 5\%$). 
    For this, a subset of patients was randomly sampled with the event-to-patients-at-risk ratio of roughly 1:3.   
    The overall survival time and status were used for survival analysis.

    Data: A .csv file containing already standardized expression data.

    \subsubsection*{GSE4922}
    
    In this dataset, two separate breast cancer cohorts can be accessed: the Uppsala (n=249) and the Singapore (n=40) data \cite{Ivshina2006a}.
    Thus, from the initial 289 samples, we selected 80 using
    the criteria mentioned above. The recurrence-free survival time and status were used for survival analysis.

    \subsubsection*{GSE7390}
    
    Gene expression data of frozen samples from 198 lymph-node negative systemically untreated patients were collected at the 
    Bordet Institute \cite{Desmedt2007a}. From these data we selected 134 using
    the criteria mentioned above. The distant-metastasis-free survival time and status were used for survival analysis.

    \subsubsection*{GSE9893}
    
    This dataset contains samples from a cohort of 132 primary tumors from tamoxifen-treated patients whose expression profiles
    were conducted at the whole genome level by 70-mer oligonucleotide microarrays containing 22,680 probes \cite{Chanrion2008a}.
	From the initial 132 samples we selected 86 using
	the criteria mentioned above. The distant-metastasis-free survival time and status were used for survival analysis.
    
    Data: Already standardized ExpressionSet.

    \subsubsection*{NKI}
    
    These are gene expression data of breast cancer tumors collected at the Netherlands Cancer Institute, as described in
    \cite{van2002gene} and \cite{van2002expression}.
    The R package \textit{BreastCancerNKI} (\url{https://bioconductor.org/packages/breastCancerNKI/}) was employed to download the data.
    From the initial 337 samples we selected 179 using
    the criteria mentioned above. The recurrence-free survival time and status were used for survival analysis.
    
    Data: Already transformed ExpressionSet from \textit{BreastCancerNKI}.

	\section*{Supplementary Reported Gene Selections}
	
	The authors of \cite{huang2018genes} collected gene lists of 33 reported signatures in breast cancer.
	They inquired PubMed for breast cancer gene signatures or classifiers, and collected the lists of gene names 
	from the original publications. As a result, they prepared gene lists containing the HUGO gene symbols.
	These lists and a detailed description of their procedure can be found in Additional Files in their publication \cite{huang2018genes}. 
	
	We downloaded these lists and adopted their procedure: not only several studies used different gene alias names, but the used gene names
	differed also across datasets. Thus, we identified all gene names, i.e. aliases, as official gene symbols individually in each dataset using the R package \textit{org.Hs.eg.db} \cite{carlson2019org}. 
	
	For all datasets, since some gene names were missed in a particular dataset, we also identified the most coexpressed genes in this dataset
	by querying COXPRESdb: a database of coexpressed genes \cite{obayashi2007coxpresdb}. Table \ref{table:gene_lists} shows 
	the resulting number of extracted genes for all datasets. In the first column, the full names of reported signatures are listed.
	We adopted the exact gene names of reported signatures from Additional file 2 in \cite{huang2018genes}.   
	The second column provides the number of genes used in the signatures. Other columns provide the number of extracted genes
	for each dataset. As mentioned in \cite{huang2018genes}, within some signatures the number of reported genes is less than the number of extracted genes, since these genes are duplicated with different probe names. For the sake of completeness, we decided
	to include all gene lists into our analysis.
	
	All signatures except the following five signatures  are used for prognosis:  The GCN of MET and HGF, 28-gene expression profile, 
	92 predictor, 85-gene signature, and 512-gene signatures are used for prediction, i.e. to predict the response to treatment or drug.

	\begin{table}[htp!]
		\caption{Gene lists from published signatures considered in the analysis.}
		\label{table:gene_lists}
		\centering
		\scalebox{0.6}{
		\begin{tabular}{lrrrrrrrrr}
			\hline
			Signature & gene \# & METABRIC & GSE11121 & GSE96058 & GSE7390 & GSE9893 & NKI & TCGA & GSE4922 \\ 
			\hline
			B-cell:IL8 ratio \cite{rody2011clinically} &  22 &   7 &  11 &   9 &  11 &   9 &   5 &  14 &  10 \\ 
			Breast cancer index \cite{ma2008five} &   7 &   7 &   7 &   7 &   7 &   7 &   7 &   7 &   7 \\ 
			Cell cycle pathway signature \cite{liu2008identification} &  26 &  26 &  26 &  26 &  26 &  26 &  26 &  26 &  25 \\ 
			92-gene predictor \cite{chang2003gene} &  92 &  80 &  80 &  80 &  80 &  80 &  80 &  80 &  78 \\ 
			EndoPredict assay \cite{filipits2011new} &   8 &   8 &   8 &   8 &   8 &   8 &   8 &   8 &   8 \\ 
			GCNs of MET and HGF \cite{minuti2012increased} &   2 &   2 &   2 &   2 &   2 &   2 &   2 &   2 &   2 \\ 
			8-gene genomic grade index \cite{toussaint2009improvement} &   4 &   4 &   4 &   4 &   4 &   4 &   4 &   4 &   4 \\ 
			97-gene genomic grade index \cite{sotiriou2006gene} &  97 &  87 &  87 &  87 &  87 &  68 &  86 &  87 &  83 \\ 
			158-gene HER2-derived prognostic predictor \cite{staaf2010identification} & 158 & 152 & 151 & 152 & 151 & 152 & 152 & 152 & 151 \\ 
			HOXB13:IL17 BR ratio \cite{ma2006hoxb13} &   2 &   2 &   2 &   2 &   2 &   2 &   2 &   2 &   2 \\ 
			186-invasivenessgene signature \cite{liu2007prognostic} & 186 & 151 & 150 & 153 & 150 & 149 & 150 & 152 & 145 \\ 
			IHC4 Score \cite{cuzick2011prognostic} &   4 &   4 &   4 &   4 &   4 &   4 &   4 &   4 &   4 \\ 
			7-gene immune response module \cite{teschendorff2007immune} &   7 &   6 &   6 &   6 &   6 &   6 &   6 &   7 &   6 \\ 
			85-gene signature \cite{iwao2005prediction} &  85 &  50 &  50 &  50 &  50 &  50 &  50 &  50 &  49 \\ 
			54-gene lung metastasis signature \cite{minn2005genes} &  54 &  54 &  54 &  54 &  54 &  54 &  54 &  54 &  52 \\ 
			MAGE-A \cite{karn2012melanoma} &   2 &   2 &   2 &   2 &   2 &   1 &   1 &   2 &   2 \\ 
			70-gene signature \cite{van2002gene} &  70 &  61 &  61 &  61 &  61 &  60 &  61 &  62 &  61 \\ 
			368-gene medullary breast cancer like signature \cite{sabatier2011gene} & 368 & 359 & 354 & 363 & 354 & 336 & 350 & 361 & 336 \\ 
			14-gene metastasis score \cite{tutt2008risk} &  14 &  14 &  14 &  14 &  14 &  14 &  14 &  14 &  13 \\ 
			Multigene HRneg/Tneg signature \cite{yau2010multigene} &  14 &  14 &  14 &  14 &  14 &  14 &  14 &  14 &  14 \\ 
			26-gene signature \cite{karn2011homogeneous} &  26 &  19 &  18 &  19 &  18 &  17 &  18 &  19 &  18 \\ 
			264-gene signature \cite{karn2011homogeneous} & 264 & 210 & 206 & 210 & 206 & 186 & 198 & 211 & 203 \\ 
			512-gene signature \cite{thuerigen2006gene} & 512 & 352 & 350 & 353 & 350 & 345 & 349 & 353 & 343 \\ 
			32-gene p53 status signature \cite{miller2005expression} &  32 &  19 &  19 &  19 &  19 &  19 &  19 &  19 &  19 \\ 
			PAM50 assay \cite{sorlie2003repeated} &  50 &  50 &  50 &  50 &  50 &  49 &  50 &  50 &  49 \\ 
			64-gene expression signature \cite{pawitan2005gene} &  64 &  48 &  48 &  48 &  48 &  46 &  48 &  48 &  47 \\ 
			127-gene classifier \cite{van2008pooling} & 127 & 123 & 123 & 124 & 123 & 112 & 123 & 124 & 118 \\ 
			21-gene signature \cite{paik2004multigene} &  16 &  16 &  16 &  16 &  16 &  16 &  16 &  16 &  16 \\ 
			26-gene stroma-derived prognostic predictor \cite{finak2008stromal} &  26 &  25 &  25 &  25 &  25 &  24 &  25 &  26 &  24 \\ 
			8-gene score \cite{sanchez20108} &   8 &   8 &   8 &   8 &   8 &   8 &   8 &   8 &   8 \\ 
			T-cell metagene \cite{rody2009t} &  50 &  46 &  46 &  48 &  46 &  46 &  47 &  46 &  44 \\ 
			28-gene expression profile \cite{vegran2009gene} &  28 &  24 &  24 &  24 &  24 &  23 &  24 &  24 &  24 \\ 
			76-gene signature \cite{wang2005gene} &  76 &  67 &  67 &  68 &  67 &  66 &  66 &  69 &  67 \\ 
			\hline
		\end{tabular}
	}
	\end{table}

	\section*{Supplementary Machine Learning Models}
	
	Various machine learning models have been adapted or developed to handle censored data. These models can be divided into 
	feature selections models as well as prognostic models. Some prognostic models already include one or more selection models 
	in the process of training. Below we provide a short description of the machine learning models used in our study. 
	The Cox proportional hazards model – while not a machine learning model – is used as baseline.
	
	The Machine Learning in R package (\textit{mlr}) \cite{bischl2016mlr} was employed to benchmark models and perform cross-validation. All results are based on 5 repeats of 5-fold cross-validation. All results below correspond to the tuned number of features: We tuned the number of features during each cross-validation fold, so that 
	a tuned (optimal) number of features was used for the eventual prediction. Other hyper-parameters are listed in Table \ref{table:survival_models}.

	\subsection*{Cox proportional hazards model}
	
	The Cox proportional hazards model can be regarded as the standard model for analyzing survival data \cite{cox1972regression}. 
	In this model, the effect of variables – also called covariates $ x_{1}, x_{2}, ..., x_{n}$ –
	on the time to an event of interest is evaluated. For example, an event might be death of the patient or relapse of the disease.
	Formally, the Cox model is expressed by the following hazard function: 
	
	\begin{equation*}
		h(t) = h_{0} \times \exp(x_{1}\beta_{1} + x_{2}\beta_{2} + ... + x_{n}\beta_{n}),
	\end{equation*}
	
	where $ \beta_{1}, \beta_{2}, ..., \beta_{n}$ of $n$ patients denote the regression coefficients, i.e. weights of the covariates: 
	the larger the coefficient the larger effect its covariate has on the prognosis of survival times. They are estimated by maximizing 
	the partial likelihood. The baseline hazard function $h(t)$ remains unspecified, since it is divided out by computing the proportional hazard.     
	
	The Cox model remains a highly robust model if applied to linearly independent data and under the assumption that the proportional
	hazard does not change over time. However, this model loses its robustness when applied to high dimensional data.

	\subsection*{Lasso, Ridge, and Elastic-Net Regressions}
	
	Since the Cox model generalizes poorly to high dimensional data, some  penalizing constraints are often used in the process
	of maximizing the partial likelihood . As a consequence, the regression coefficients shrink toward zero, their variances 
	reduce as well, and the less important covariates tend to have less effect in the model.     
	
	L1 and L2 regularizations are two standard forms of regularization:
	
	\begin{equation*}
	\begin{aligned}
	L1 = \lambda \times (|\beta_{1}| + |\beta_{2}| + ... + |\beta_{n}|), \\
	L2 = \lambda \times (\beta_{1}^{2} + \beta_{2}^{2} + ... + \beta_{n}^{2}),
    \end{aligned}
	\end{equation*}

	where $\lambda$ is the regularization constant. 
	The L1 regularization is also known as LASSO regression and produces models with a smaller set of coefficients, since several 
	coefficients are completely reduced to zero. Thus, variable selection is also performed during the fitting process.
	
	The L2 regularization is also known as ridge regression and shrinks all coefficients by the same factor. As a results, all coefficients 
	are reduced but none is eliminated. 
	
	The L1 regularization cannot select more variable than the number of samples. Moreover, it is biased toward the selection of groups of correlated variables \cite{zou2005regularization}. 
	
	To overcome these limitations, one can use a linear combination of L1 and L2 penalties, which is then called the \textbf{elastic net 
	regression}. The elastic net regression is especially useful when the number of variables is larger than the number of samples \cite{zou2005regularization}.   
	In our study, we evaluated the extensions of the lasso, ridge, and elastic net regressions to the Cox model \cite{tibshirani1997lasso, simon2011regularization}.
	
	\subsection*{Boosted models}

	Boosting is an ensemble learning technique that combines the so-called weak i.e. base learners into a stronger learner that are trained
	sequentially \cite{schapire1990strength}. During each iteration, a new model is added to the ensemble correcting the errors of
	the previous model. Boosting has been adapted to survival analysis \cite{friedman2000additive, hothorn2006survival}.   
	
	We used the gradient boosting in our study \cite{friedman2002stochastic}. This type of boosting trains 
	on the residual errors (gradients) of the entire ensemble model at each step. It can also be trained 
	with linear models as base learners as well as with decision trees as base learners. In this study, both methods
	were assessed.
	
	\subsection*{Survival Trees and Random Survival Forests}	
	
	Survival trees \cite{gordon1985tree} and random survival forests \cite{ishwaran2008random} are an extension of 
	decision trees \cite{morgan1963problems, breiman2017classification} and the random forests algorithm developed by 
	Leo Breiman \cite{breiman2001random} to censored survival data. 
	Decision trees and random forests, in turn,  are nonparametric regression and classification methods that are well suited 
	for the case, where the number of variables is greater than the number of samples, for example in genetics.     
	
	\subsubsection*{Survival Trees}	
	
	In general description of decision trees, the space spanned by predictor variables, 
	i.e. by the covariates, is recursively partitioned into several groups such that observations with similar responses are 
	grouped together. In the case of numeric variables such as gene expression values, binary splits are conducted. 
	For the splitting a variable and selecting a splitting threshold, decision trees follow the principle of impurity reduction.
	Following this principle, each split in the tree results in daughter nodes whose impurity is reduced in comparison to the parent nodes.
	The impurity can be measured with the Shannon entropy or the Gini index or other statistics. 
	Finally, in an ensemble of trees predictions are made by means of averaging and combining the results of each decision tree.       
	
	\subsubsection*{Random Forests}	
	
	Random survival forests aggregate the results from ensembles of decision trees, whereas each tree is generated from a bootstrap sample of the data. At each node, a random subset of predictor variables is sampled and one variable is selected to split on. 
	The selected variable maximizes the difference in survival between daughter nodes. Mathematically, the log-rank statistic 
	over all available split points and variables is maximized. For prediction, an average over the predictions of the single 
	trees is used (a vote is used for a classification problem).
	
	Both survival trees and random survival are able to robustly handle high-dimensional non-linear data and detect interactions
	among them. Provided the depth of trees is chosen carefully, they also reduce the tendency of overfitting the data.
	However, both algorithms are biased towards selecting more heterogeneous variables, i.e. variables with many possible split points are preferred. To overcome this problem, one can use the conditional inference forests \cite{hothorn2006unbiased} that 
	select the split points based on linear rank statistics. Nonetheless, to detect non-linear effects in the predictor 
	variable space,  selecting the split points using maximally selected rank statistics can be conducted \cite{lausen1992maximally}. 
	
	Alongside survival trees, we evaluated both the standard random survival forests and the maximally selected rank statistics random survival forests.
	
	The full names of the prognostic models, respective hyper-parameters as well as packages and functions used in this study are shown in
	Table \ref{table:survival_models}.
	
	\begin{table}[htp!]
		\caption{Machine learning prognostic models and respective hyper-parameters used in this study.}
		\label{table:survival_models}
		\scalebox{0.7}{
		\begin{tabular}{l|l|l|l}
			\textbf{Survival model} & \textbf{Full name of the survival model}                   & \textbf{Package and function}    & \textbf{Hyper-parameters}                                                                                                                                                              \\ \toprule
			CoxPH          & Cox proportional hazards model                    & survival, coxph        &                                                                                                                                                                                                \\ \hline
			Lasso          & Lasso regression                                  & glmnet, cv.glmnet      & alpha = 1, nfolds = 5                                                                                                                                                                          \\ \hline
			Ridge          & Ridge regression                                  & glmnet, cv.glmnet      & alpha = 0, nfolds = 5                                                                                                                                                                          \\ \hline
			Elastic-Net    & Elastic net survival regression                   & glmnet, cv.glmnet      & alpha = 0.5, nfolds = 5                                                                                                                                                                        \\ \hline
			GB-Linear      & Gradient boosting with linear learners            & mboost, gamboost                        &  baselearner =
			"bols"                                                                                                                                                                                               \\ \hline
			GB-Tree        & Gradientboosting with tree-based learners         & mboost, gamboost                       & baselearner =
			"btree"                                                                                                                                                                                               \\ \hline
			RSF            & Random survival forests                           & randomForestSRC, rfsrc & \begin{tabular}[c]{@{}l@{}}mtry: from (number of genes)/3 to 100\\ nodesize: 5 to 30, ntree=500\end{tabular}                                                                    \\ \hline
			Rank-RF        & Maximally selected rank statistics random forests & ranger, ranger         & \begin{tabular}[c]{@{}l@{}}splitrule = "maxstat",\\ importance = "permutation"\end{tabular} \\ \hline
			STree          & Survival trees                                    & rpart, rpart                        &                                                                                                                                                                                                \\ 
		\end{tabular}
	}
	\end{table}

	\section*{Supplementary Gene Selection Methods}
	
	During variable selection a subset of for the survival outcome relevant variables is selected. 
	We applied 5 different gene selection methods and measured the performance of survival models described above.
	
	In a univariate model, a univariable Cox proportional hazards model is fitted to expression values of each gene
	and the genes are ranked by the resulting C-index of the corresponding model. 
	
	The variable importance of the random forests algorithm is computed permuting the column containing the expression
	values of each gene and calculating the difference between the performances of the survival model before and after permutation. 
	Subsequently, the genes are ranked based on these differences.

	The survival random forests variable hunting method, on the other hand, use a different importance score.
	First, the standard variable importance is conducted on the entire dataset.  
	Second, a random subset of genes is selected with probability proportional to the calculated variable importances, and a forest is fitted.  
	Third, the selected genes are ordered by the shortest distance from the tree root to the largest subtree including this gene as its root; they are added successively to the fitting model until the joint importance does not increase anymore.
	These steps are iterated a specified number of times. Eventually, the variable importances result from the ranking
	of the variables based on the frequency of occurrence in these iterations. 
	
	The Minimum Redundancy Maximum Relevance algorithm \cite{peng2005feature} selects variables that are mutually far away from each other:
	variables that are mutually close to each other might be redundant. Thus, the algorithm minimizes redundancy by removing the potentially redundant variables. At the same time, the selected variables are highly correlated with the response variable such as survival time, 
	meaning that they exhibit maximum relevance. 
	
	The Conditional Variable Importance for Random Forests utilizes the linear rank statistics of conditional random forests 
	described above. 
	
	The full names of the selection methods, respective hyper-parameters as well as packages and functions used in this study are shown in
	Table \ref{table:selection_models}.
	
		\begin{table}[htp!]
		\caption{Machine learning selection methods and respective hyper-parameters used in this study.}
		\label{table:selection_models}
		\scalebox{0.6}{
		\begin{tabular}{l|l|l|l}
			\textbf{Selection method} & \textbf{Full name of the selection method}                   & \textbf{Package and function}    & \textbf{Hyper-parameters}                                                                                                                                                                        \\ \toprule
			CF     & Conditional variable importance for random forests & party, varimp               & conditional = TRUE                                                                                                      \\ \hline
			SRC    & Random survival forests with variable importance   & randomForestSRC, rfsrc      & \begin{tabular}[c]{@{}l@{}}ntree = 500,  nsplit = 10\\ mtry = (number of features)/3, nodesize=5\end{tabular}           \\ \hline
			SRC-VH & Random survival forests with variable hunting      & randomForestSRC, var.select & \begin{tabular}[c]{@{}l@{}}method = "vh", ntree=500, \\ nodesize=5, splitrule="logrank", \\ nsplit=10, K=5\end{tabular} \\ \hline
			MRMR   & Minimum redundancy maximum relevance filter        & mRMRe, mrmr                 &                                                                                                                         \\ \hline
			UM     & Univariate model                                   & mlr, various                &                                                                                                                         \\ \hline
		\end{tabular}
	}
	\end{table}

	\section*{Supplementary Random Signature Size}
	
	According to the rule of thumb that Cox proportional hazards models should be used with a minimum of 10 events per predictor variable (EPV), we should use 2 to 16 random genes in our datasets, since TCGA and METABRIC contain the smallest and largest numbers of 17 and 164 events, respectively. 
	However, this rule is based on two simulation studies and may be relaxed \cite{vittinghoff2006}. Moreover, a study investigating 
	this rule of thumb in 2 million anonymized patient records suggested that sample size for developing prognostic models is not simply related to EPV and that EPV should be dataset dependent. 
	
\begin{figure}[htp!]
		\centering
		\includegraphics[width=0.8\textwidth]{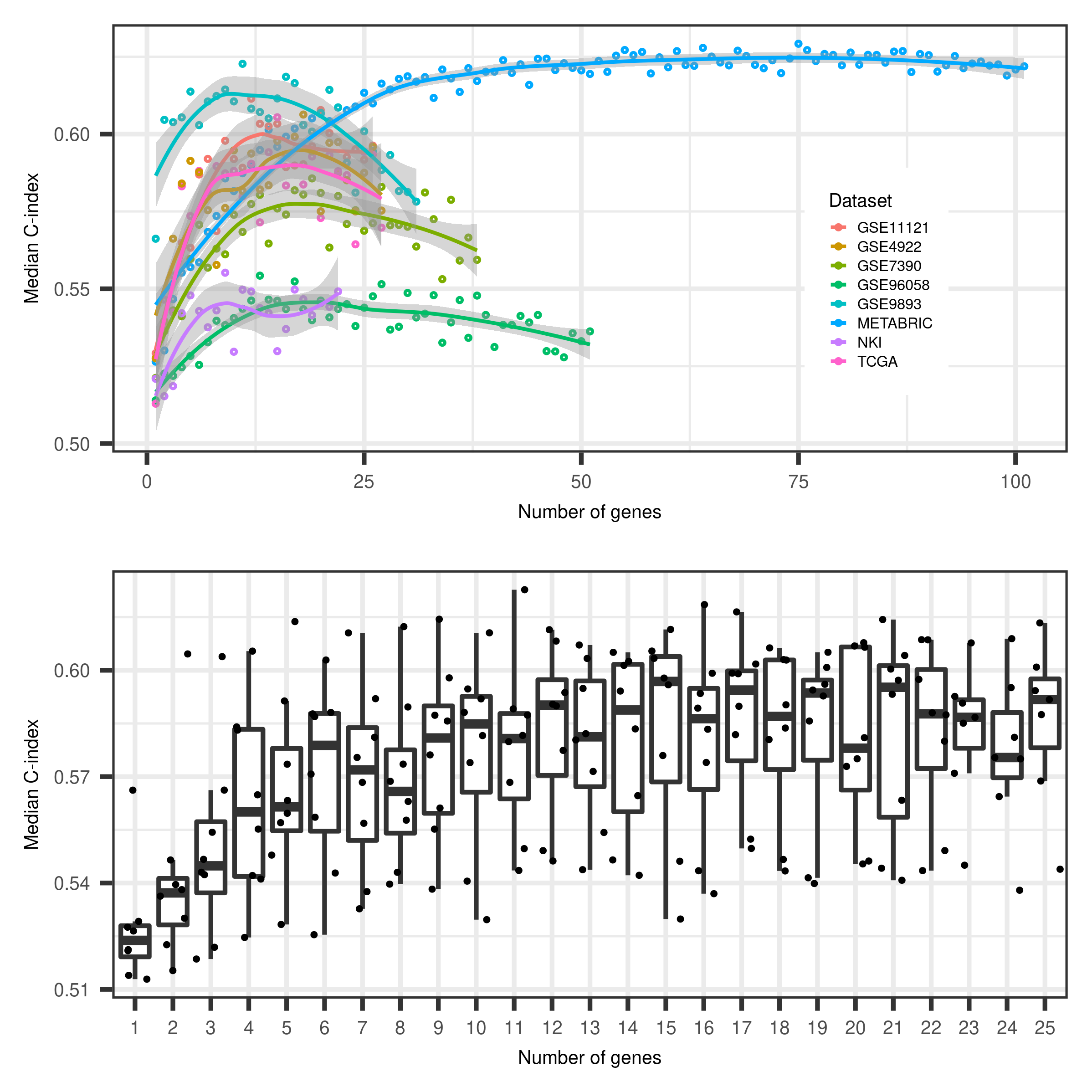}
		\caption{\textbf{Relationship of median C-index and the number of genes in a random gene set.}
		The median C-index was calculated based on 100 genes set sampled at random for each data point.
		The Cox proportional hazards model was used as prognostic model. (top) Median C-indices computed 
		based on individual dataset. (bottom) Distribution of C-indices for all datasets in the range from 1-20 genes 
		in a gene set.   
	    }
		\label{fig:signature_size}
	\end{figure}

	For these reasons, we randomly sampled gene sets containing different number of genes ranging from 1-101 for each datasets. 
	The sampling was repeated 100 times for each number of genes, Cox models were fitted, and the median C-index was calculated. 
	As can be seen in the top Figure \ref{fig:signature_size}, the prognostic power is indeed dataset dependent. For all datasets except METABRIC, we could investigate a limited number of genes, since the Cox model does not converge with a smaller number of events.
	Nonetheless, we see that prognostic power reaches a plateau if a sufficient number of events is considered (METABRIC).
	For the rest datasets, the optimal number of genes seems to be in the range of 15-25 genes. In the bottom Figure  \ref{fig:signature_size}, this
	range can be inspected more closely. As can be seen, the distribution of the median C-indices increases in the 
	range from 1-15 genes in a gene set, after which it seems to fluctuate around a constant value.
	Chou et. al. have shown that the optimal number of genes in a signature lies around 20 and that with a larger number of genes
	a model tends to overfit data (Figure 4 in \cite{chou2013gene}). Moreover, most clinically relevant gene-expression signatures tend to contain a smaller number of genes varying from 2-50 (\textit{Supplementary Reported Gene Selections}). Thus, we chose to sample 20 random genes in all datasets.                        
	
	\section*{Supplementary Random Signature Superiority}
	
	In order to examine whether Random Signature Superiority (RSS) is present in this study, we 
	we calculated the number of random signatures performing above the C-index of 
	the reported 26-gene signature \cite{karn2011homogeneous} (which has roughly the same size as random signatures, see Supplementary Reported Gene Selections) for each prognostic model and averaged this value over all datasets.
	The results are shown in Figure \ref{table:rss}. 
	We found (\textit{Supplementary Random Signature Superiority}) that more than 60\% of random signatures outperform the aforementioned reported signature in 4 of 8 datasets, exactly 49\% in one dataset, and less than 22\% in the remaining 3 datasets. Averaging across datasets, 
	44\% of random signatures outperform the aforementioned reported signature.

		\begin{table}[htp!]
		\centering
	\caption{Evaluation of the Random Signature Superiority.}
	\label{table:rss}	
	\begin{tabular}{l|r}
		\textbf{Dataset} & \textbf{Average [\%]}\\
		\toprule
		GSE11121 & 62\\
		\hline
		GSE4922 & 67\\
		\hline
		GSE7390 & 8\\
		\hline
		GSE96058 & 22\\
		\hline
		GSE9893 & 49\\
		\hline
		METABRIC & 64\\
		\hline
		NKI & 19\\
		\hline
		TCGA & 60\\
		\midrule
		All datasets & 49 \\
	\end{tabular}
	\end{table}

	\section*{Supplementary Difference Between Selection Methods}
	
\begin{figure}[ht]
		\centering
		\includegraphics[width=\textwidth]{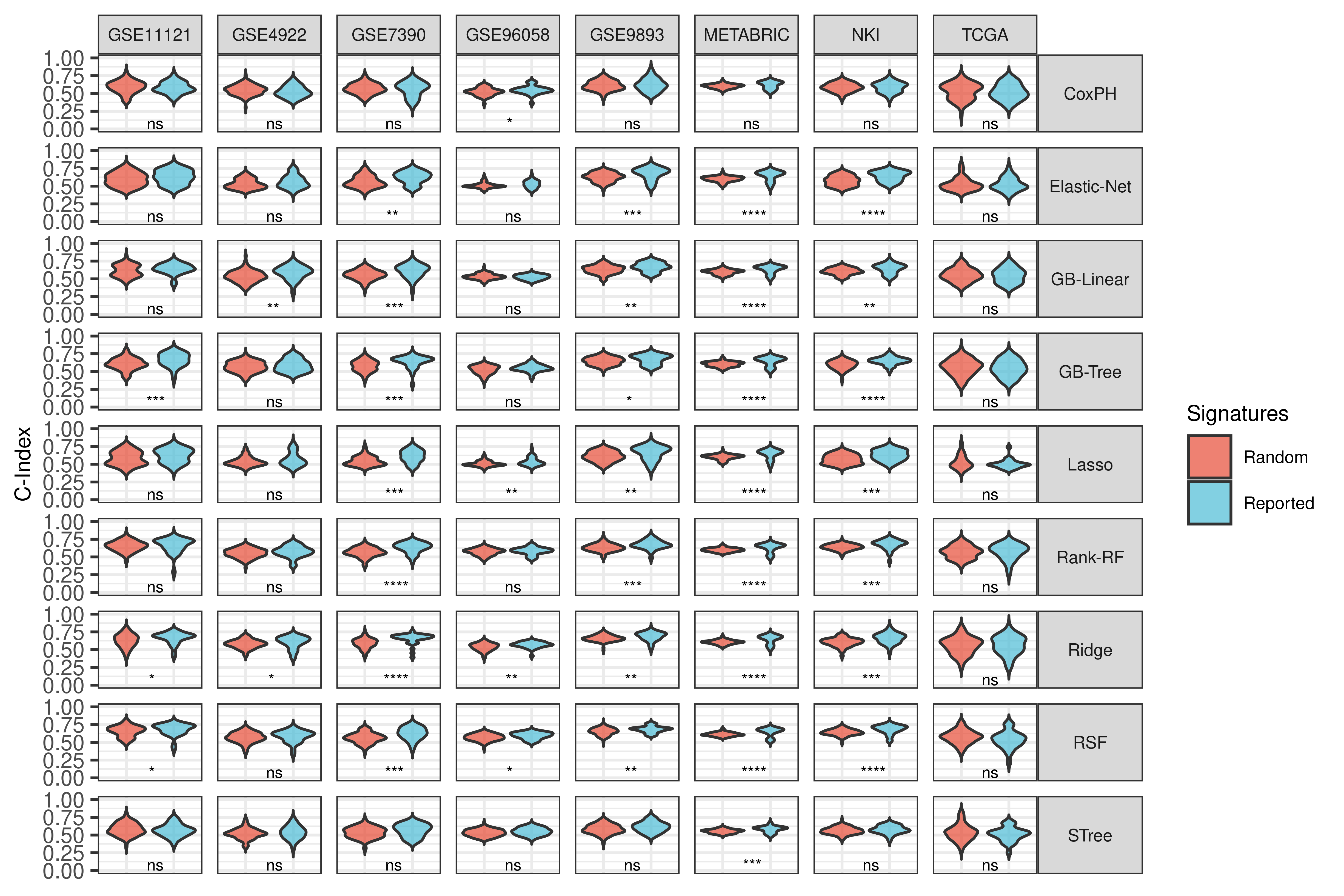}
		\caption{\textbf{Comparison of C-indices for random and reported selection methods.}}
		\label{fig:diff_rand_reported}
	\end{figure}

	In order to inspect the differences in prognostic power between random and reported selection methods, 
	we plot the distributions  in form of the violin plots for each model and each dataset in Figure \ref{fig:diff_rand_reported}. 
	The distributions are compared using the Wilcoxon rank sum test. The according significance levels are shown below violin plots.
	As can be inspected here, reported signatures tend to have higher C-indices than random signatures, although the level of statistical significance varies across models and datasets.
	
	\section*{Supplementary Dataset Dependency}
	
\begin{figure}[htp!]
		\centering
		\includegraphics[width=\textwidth]{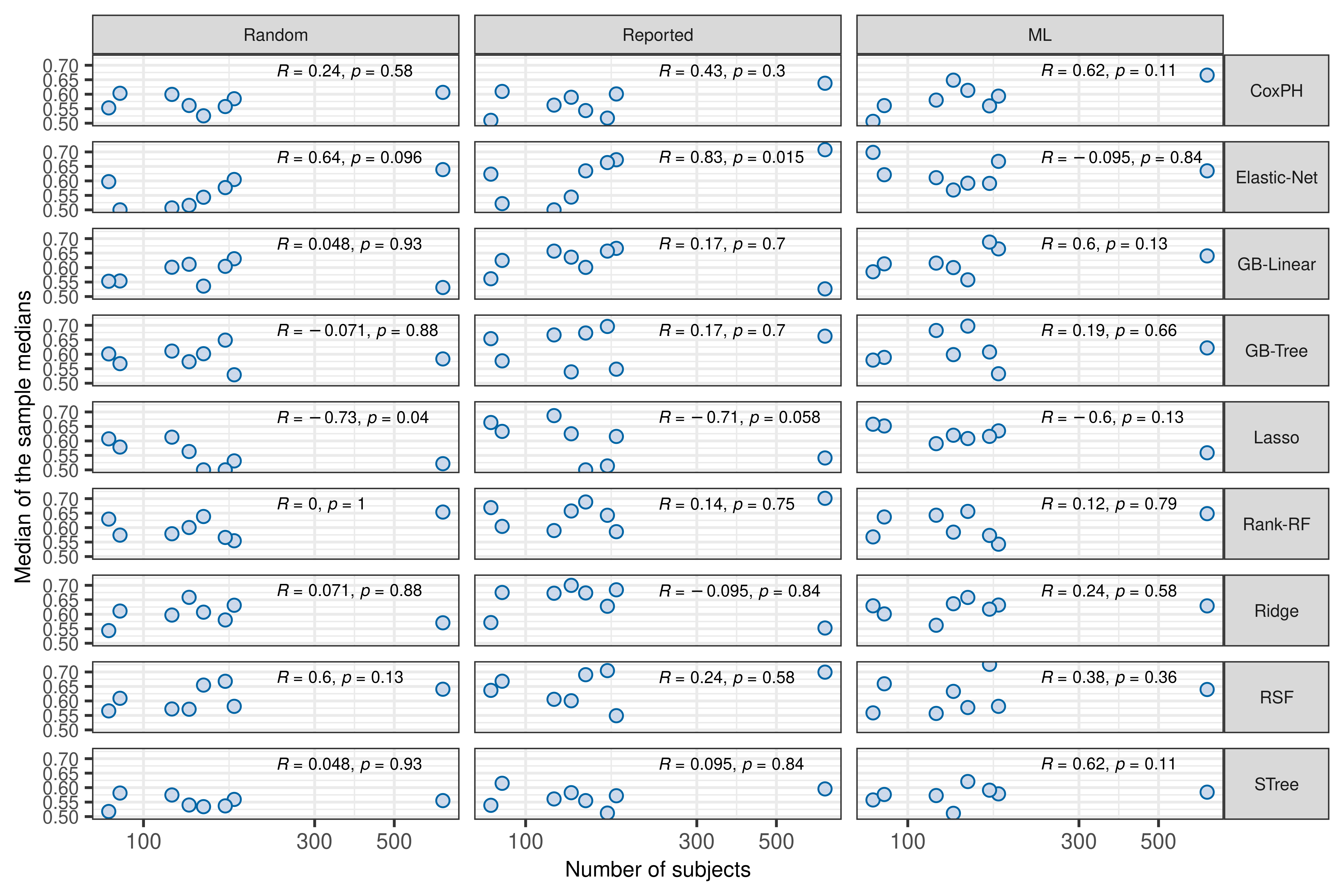}
		\caption{\textbf{Median of the sample medians in relation to the number of subjects.}}
		\label{fig:mom_nop}
	\end{figure}

\begin{figure}[htp!]
	\centering
	\includegraphics[width=\textwidth]{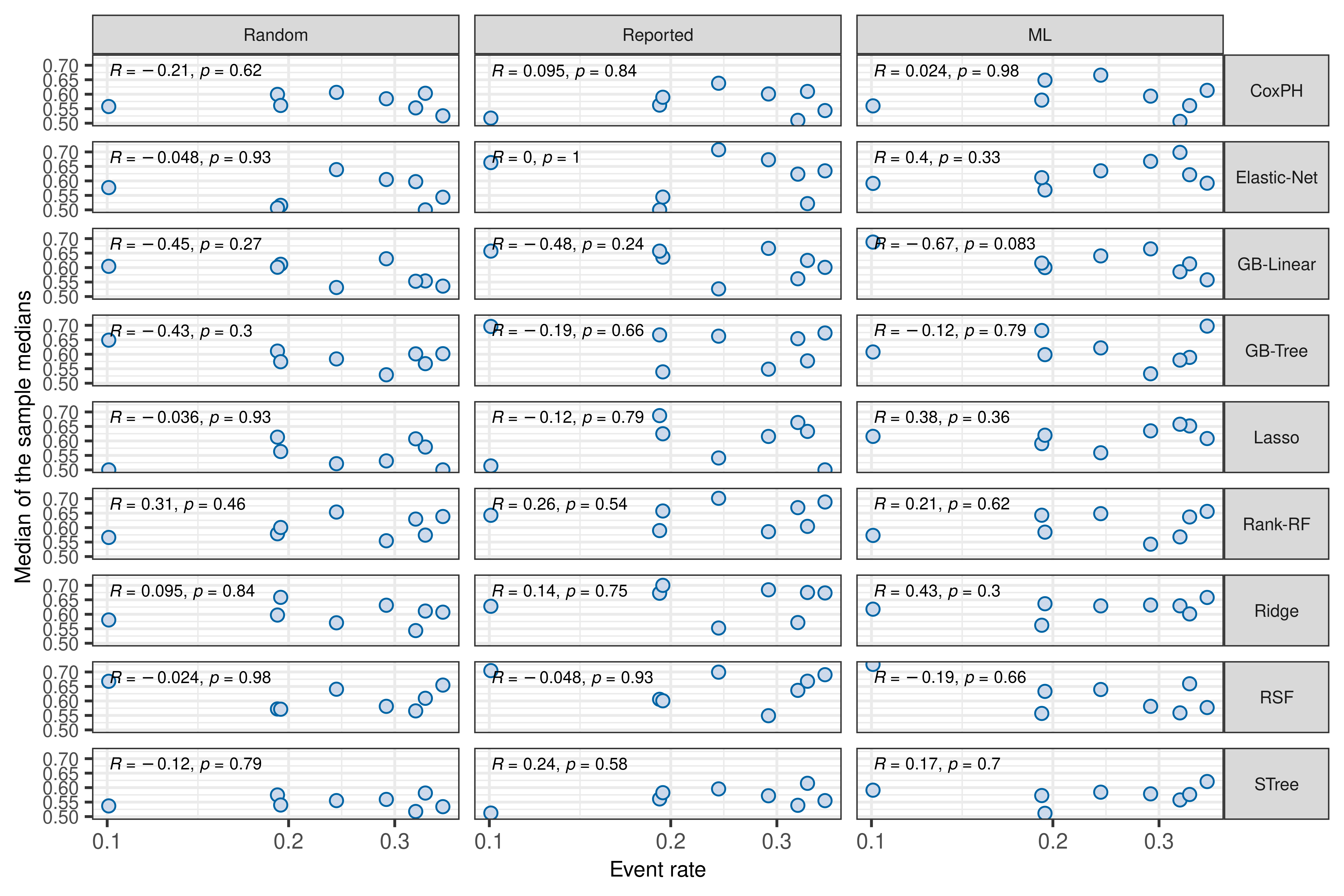}
	\caption{\textbf{Median of the sample medians in relation to the event rate.}}
	\label{fig:mom_er}
    \end{figure}

\begin{figure}[htp!]
	\centering
	\includegraphics[width=\textwidth]{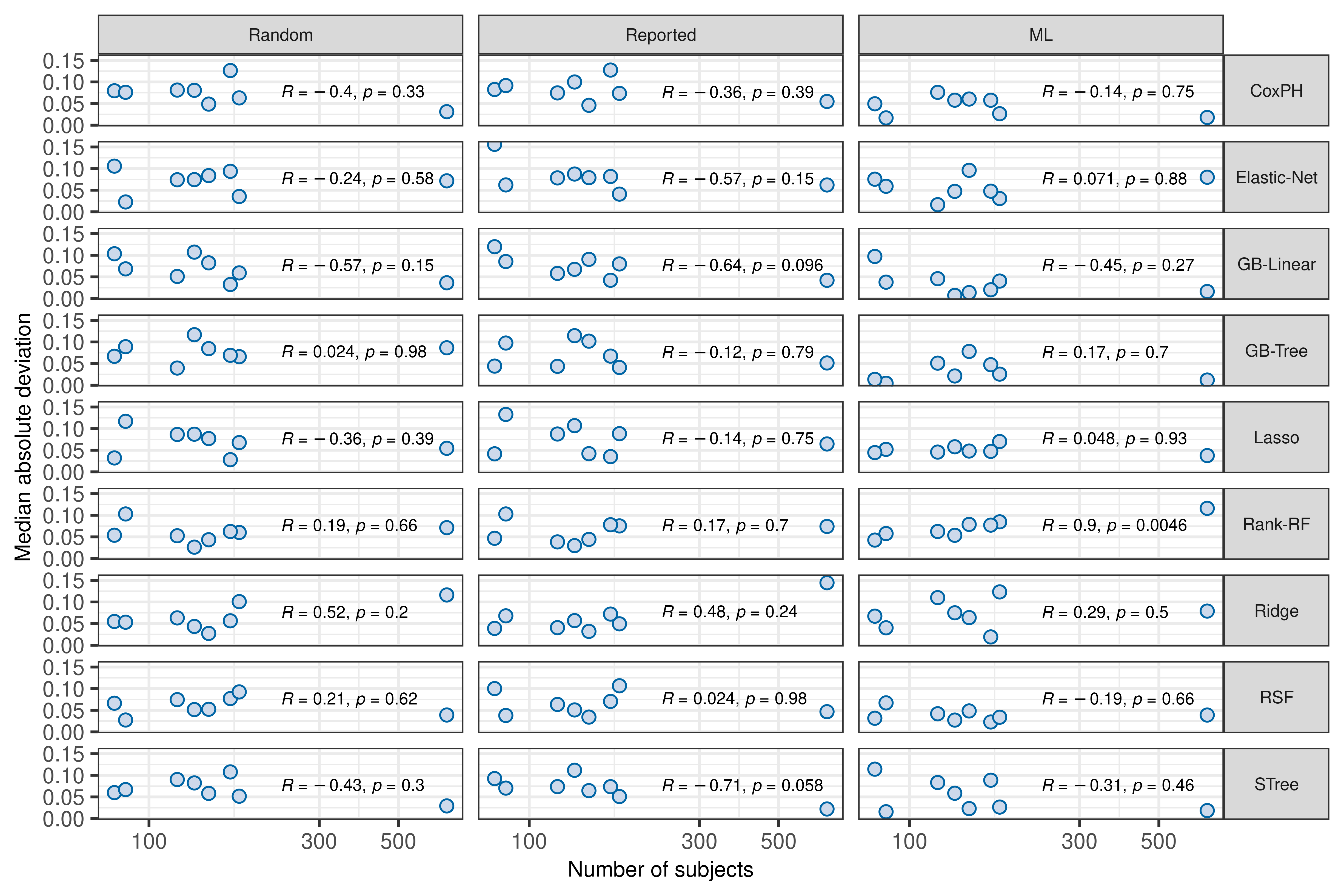}
	\caption{\textbf{Median absolute deviation in relation to the number of subjects.}}
	\label{fig:mad_nop}
	\end{figure}

\begin{figure}[htp!]
		\centering
		\includegraphics[width=\textwidth]{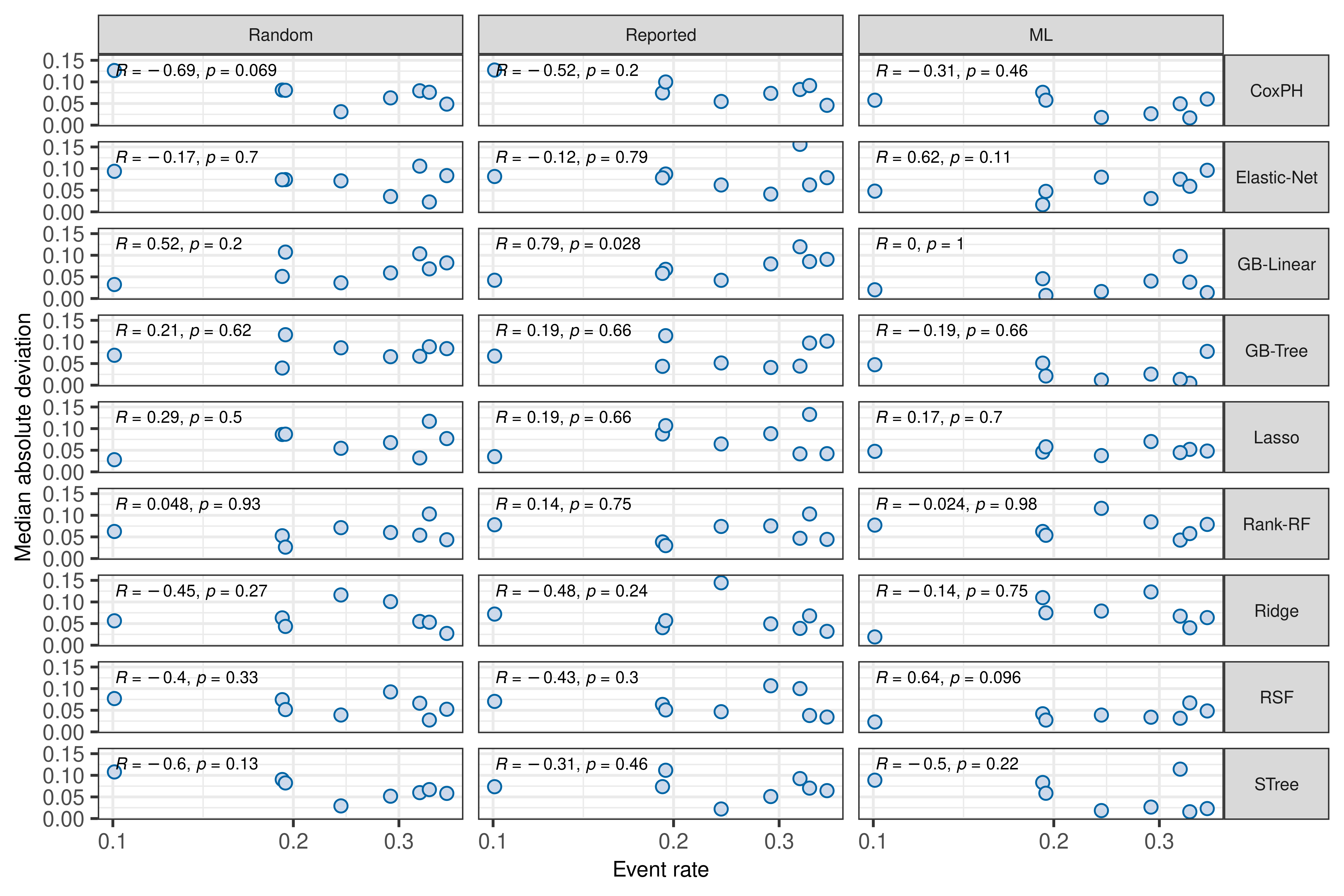}
		\caption{\textbf{Median absolute deviation in relation to the event rate.}}
		\label{fig:mad_er}
	\end{figure}

	We investigated whether the median of the sample medians (MOM) and the median absolute 
	deviation (MAD) correlate with the number of subjects 
	as well as with the event rate in a dataset.        
	Figures \ref{fig:mom_nop}, \ref{fig:mom_er}, \ref{fig:mad_nop}, \ref{fig:mad_er} plot the results for each prognostic model along with the corresponding Spearman's rank correlation coefficients and their p-values. As can be inspected in these plots, the MOM and MAD seem to be uncorrelated with both the number of subjects and with the event rate.  
	
	\section*{Supplementary Combined Dataset}
	
\begin{figure}[htp!]
		\centering
		\includegraphics[width=\textwidth]{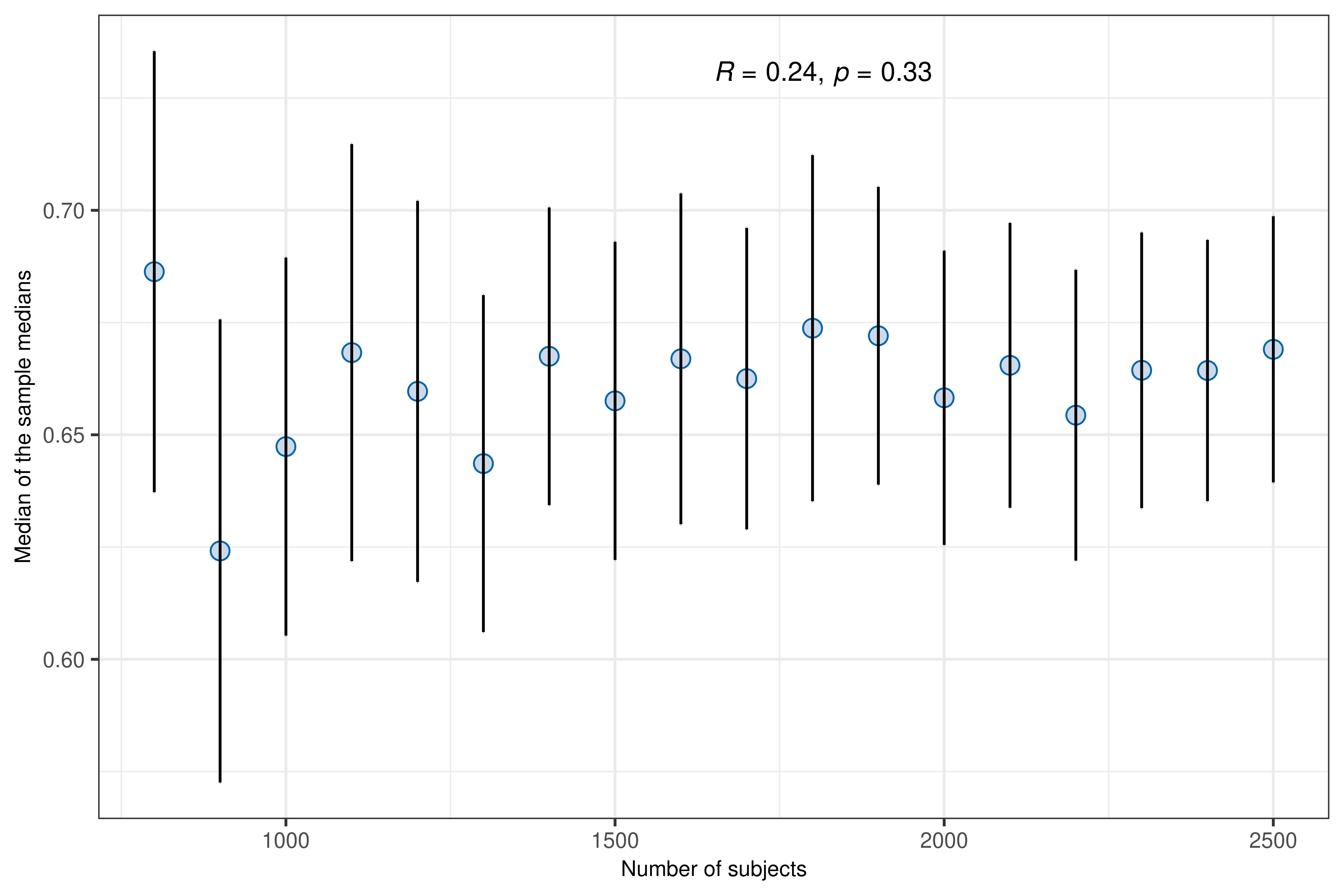}
		\caption{\textbf{Median of the sample medians (MOM) in relation to the number of subjects.}
		Each data points represents the MOM from 1000 samples, whereby 20 genes are selected at random and the
	    dataset identification is included as covariate in the Cox proportional hazards models. The black lines indicate 95\% confidence intervals for the MOM.}
		\label{fig:mad_combined}
	\end{figure}

	In order to investigate large sample sizes, we 
	combined the 8 datasets into one large dataset resulting in 2553 subjects (683+86+134+1102+120+179+80+169 = 2553).
	As can be seen from the number of subjects, all 1102 subjects  from the GSE96058 dataset were integrated without downsampling,
	since the resulting event rate of 0.15 seemed to be sufficient. The datasets were standardized, normalized, and annotated as described in section \textit{Supplementary Data}.
	A list of genes common to all datasets (3969 genes) was extracted in order to combine the datasets based on this list. 
	Z-score transformation was applied to a single dataset \cite{cheadle2003analysis}. Subsequently, the single datasets were combined into one large dataset. 
	The following sampling procedure was applied: 20 genes were selected at random, the dataset identification was included as 
	covariate in the Cox proportional hazards model in order to directly correct for batch effects, and the median C-index was measured.
	This sampling procedure was repeated 1000 times – resulting in 1000 different random signatures – and the median of the sample medians was computed.   
	We resampled the data with different sample sizes ranging from 800 to 2500 subjects. We kept the event rate constant (event rate = 0.15) in each sample in order to investigate the relation between sample size and prognostic performance, since the we have already 
	shown (\textit{Supplementary Dataset Dependency}) that larger event rates do not increase prognostic power.    
	As can be seen in Figure \ref{fig:mad_combined},
	the performance does not increase with larger sample sizes (Pearson correlation coefficient $R = 0.24, p = 0.33$).   
	
	\section*{Supplementary Event Type}
	
\begin{figure}[htp!]
	\centering
	\includegraphics[width=\textwidth]{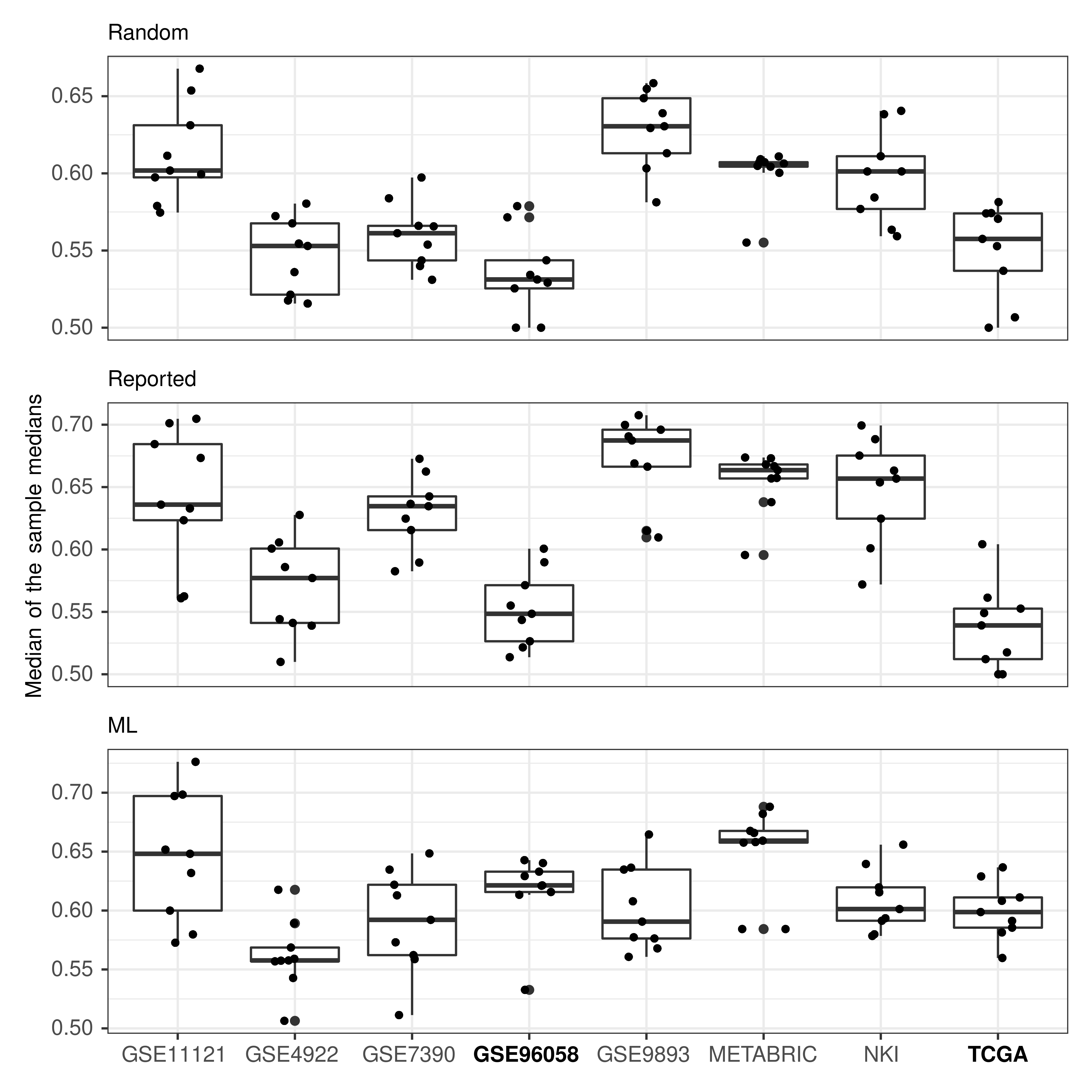}
	\caption{\textbf{Median of the sample medians (MOM) in relation to datasets.}
		Each data points represents the MOM based on C-indices for each selection model. Both datasets
	GSE96058 and TCGA – where the overall survival
	was used for the prognosis – are shown in bold. (Random) Signatures selected at random. (Reported) Reported signatures from the literature. (ML) Signatures selected with machine learning.}
	\label{fig:event_type}
\end{figure}

	In Fig. \ref{fig:event_type} box plots of median C-indices are shown across datasets. Each data point represents
	the MOM per selection method (9 data points corresponding to 9 selection methods described above). Only in GSE96058 and
	TCGA the overall survival was used for the prediction of survival times, which may include events that are not related to the disease.The prediction of disease-free, distant-metastasis-free, and recurrence-free survival is more specific than
	the prediction of the overall survival and was used in other datasets. As can be seen in Fig. \ref{fig:event_type},
	both datasets show the lowest MOMs for the reported signatures and random signatures, although the MOM of TCGA is comparable with the MOMs of GSE4922 and GSE7390. In regards to signatures based on machine learning (ML), on would expect that the ML models are trained to select genes according to the specific event type provided in a dataset, so that no substantial difference based on event type should be observed across datasets, which can be seen in Fig. \ref{fig:event_type} (ML).     
	
\newpage

\bibliography{SI.bib}
\bibliographystyle{Science}